\documentclass[12pt]{article}
\usepackage{amsfonts}
\usepackage{amsmath}
\newcommand{\R}{\mathbb R}

\newcommand{\rowspace}{\rule{0pt}{16pt}}

         \usepackage{graphics}
\begin{document}
%\begin{frontmatter}

\title{Where Did The Moon Come From?} 
\author{Edward Belbruno\footnote{Program in Applied and Computational
Mathematics, Princeton University, Princeton, NJ 08544
(belbruno@math.princeton.
edu)} \hspace{.00in} and 
J. Richard Gott III\footnote{Department of Astrophysical Sciences,
Princeton
University, Princeton, NJ 08544 (jrg@astro,princeton.edu)}\\
}
\date{January 6, 2005}
\maketitle
\begin{abstract}
The current standard theory of the origin of the Moon is that the
Earth was hit by a giant impactor the size of Mars causing ejection
of iron poor impactor mantle debris that coalesced to form the Moon.
But where did this Mars-sized impactor come from? Isotopic evidence
suggests that it came from 1AU radius in the solar nebula and computer
simulations are consistent with it approaching Earth on a zero-energy
parabolic trajectory. But how could such a large object form in the disk
of planetesimals at 1AU without colliding with the Earth early-on before
having a chance to grow large or before its or the Earth's iron core had
formed? We propose that the giant impactor could have formed in a stable
orbit among debris at the Earth's Lagrange point $L_4$ (or $L_5$). We
show such a configuration is stable, even for a Mars-sized impactor. It
could grow gradually by accretion at  $L_4$ (or $L_5$), but eventually
gravitational interactions with other growing planetesimals could kick
it out into a chaotic creeping orbit which we show would likely cause
it to hit the Earth on a zero-energy parabolic trajectory. This paper
argues that this scenario is possible and should be further studied.

\end{abstract}
%\end{frontmatter}

\section{Introduction}
\label{Section:1} 

The currently favored theory for the formation of the Moon is the giant
impactor theory formulated by Hartmann $\&$ Davis (1975) and Cameron
$\&$ Ward (1976). Computer simulations show that a Mars-sized giant impactor
could have hit the Earth on a zero-energy parabolic trajectory, ejecting impactor
mantle debris that coalesced to form the Moon. Further studies of this
theory include (Benz, Slattery, $\&$ Cameron 1986, 1987; Benz,
Cameron, $\&$ Melosh 1989; Cameron $\&$ Benz 1991; Canup $\&$
Asphaug 2001; Cameron 2001, Canup 2004; Stevenson 1987). 
We summarize
evidence favoring this theory: (1)
It explains the lack of a large iron core in the Moon. By the late time
that the impact had taken place, the iron in the Earth and the giant impactor had already sunk
into their cores. So, when the Mars-sized giant impactor hit the Earth in a
glancing blow, it expelled debris, poor in iron, primarily from mantle of the giant
impactor which eventually
coalesced to form the Moon (cf. Canup 2004, Canup2004B). Computer simulations (assuming a zero-energy
parabolic trajectory for the impactor) show that iron in the core of
the giant impactor melts and ends up deposited in the Earth's core. (2)
It explains the low (3.3 grams/cm$^3$) density of the Moon relative to
the Earth (5.5 grams/cm$^3$), again due to the lack of iron in the Moon.  (3) It
explains why the Earth and the Moon have the same oxygen isotope abundance
- the Earth and the giant impactor came from the same radius in the solar
nebula. Meteorites originating from the parent bodies of Mars and Vesta, from different neighborhoods in the solar
nebula have different oxygen isotope abundances. The impactor theory is able to
explain the otherwise paradoxical similarity between the oxygen isotope
abundance in the Earth combined with the difference in iron. This is
perhaps its most persuasive point. (4) It explains, because it is due to
a somewhat unusual event, why most planets (like Venus and Mars, Jupiter
and Saturn) are singletons, without a large moon like the Earth. Competing
ideas have not had comparable success. For example,
the idea that the Earth and the Moon formed together as sister planets
in the same neighborhood fails because it doesn't explain the difference
in iron. Whereas the idea that the Moon formed elsewhere in the solar
nebula and was captured into an orbit around the Earth fails because its
oxygen isotope abundances would have to be different. That a rapidly
spinning Earth could have spun off the Moon(from mantle material) is
not supported by energy and angular momentum considerations, it is argued.

Still, the giant attractor theory has some puzzling aspects. Planets
are supposed to grow from planetesimals by accretion. How did an object
so large as Mars, form in the solar nebula at exactly the same radial
distance from the Sun without having collided with the Earth earlier,
before it could have grown so large. Indeed, such must have been the
case during the formation of Venus and Mars for example. It's also
hard to imagine an object as large as Mars forming in an eccentric
Earth-crossing orbit. One might expect large objects forming in the
solar nebula to naturally have nearly circular orbits in the ecliptic plane, like the Earth
and Venus.  Besides, a Mars-sized object in an
eccentric orbit would not be expected to have identical oxygen abundances
relative to the Earth, and would collide with the Earth on a hyperbolic
trajectory not the parabolic trajectory that the successful computer simulations of the
great impact theory have been using.(Recent collision simulations by Canup(2004) place an upper
limit of 4 km/s for the impactor's velocity-at-infinity approaching the Earth, 
setting an upper limit on its eccentricity of $\stackrel{<}{\sim} 0.13$.) The
Mars-sized object needs to form in a circular orbit of radius 1 AU in the
solar nebula but curiously must have avoided collision with the Earth
for long enough for its iron to have settled into its core. Is
there such a place to form this Mars-sized object?

Yes: the Earth's Lagrange point $L_4$(or$L_5$) which is at a
radius of 1 AU from the Sun, with a circular orbit $60^o$ behind the
Earth(or $60^o$ ahead of the Earth for $L_5$). 
After the epoch of gaseous dissipation in the inner solar nebula has passed we are left with a
thin disk of planetesimals interacting under gravity.
The three-body
problem shows us that the Lagrange point $L_4$(or equivalently $L_5$) for the Earth
is stable for a body of negligible mass even though it is maximum in the effective 
potential. Thus, planetesimals can be trapped near $L_4$ and as they are perturbed they will
move in orbits that can remain near this location.  This remains true as the Earth grows by
accretion of small planetesimals. Therefore, over time, it might not be surprising
to see a giant impactor growing up at $L_4$(or $L_5$). In the
Discussion section we argue that there are difficulties in having the giant impactor
come from a location different from $L_4$(or $L_5$).

Examples of planetsimals remaining at Lagrange points of other bodies include the 
well known Trojan asteroids
at Jupiter's $L_4$ and $L_5$ points. As another
example, asteroid 5261 Eureka has been discovered at Mars' $L_5$
point. (There are five additional asteroids also thought to be Mars Trojans: 1998 $VF_{31}$, 1999 $UJ_7$, 2001 $DH_{47}$, 2001 $FG_{24}$,
2001 $FR_{127}$.) The Saturn system has several examples of bodies existing at the equilateral Lagrange points of 
several moons, which we discuss further in the "Note added" after the Discussion section.  

We propose that the Mars-sized giant impactor can form as part of debris
at Earth's $L_4$ Lagrange point. (It could equally well form at
$L_5$, but as the situation is symmetric, we will simply refer in the rest
of the paper, unless otherwise indicated, to the object forming at $L_4$;
the argument being the same in both cases). As the object forms and gains
mass at $L_4$, we can demonstrate that its orbit about the Sun remains stable.
Thus, it has a stable orbit about the Sun, and remaining at $L_4$ keeps
it from collision with the Earth as it grows.  Furthermore, this orbit
is at exactly the same radius in the solar nebula as the Earth so that
its oxygen isotope abundances should be identical. It is allowed to
gradually grow and there is time for its iron to settle into its core,
and the same also happens with the Earth.  The configuration is stable
providing the mass of the Earth and the mass of the giant impactor are
both below .0385 of the mass of the Sun, which is the case. But eventually, we numerically demonstrate that 
gravitational perturbations from 
other growing planetesimals can kick the giant impactor into a horseshoe orbit
and finally into an orbit which is chaotically unstable in nature allowing escape from
$L_4$. The giant impactor can then enter an orbit about the Sun 
which is at an approximate radial distance
of 1 AU, which will gradually creep toward the Earth; leading, with
large probability, to a nearly zero-energy  parabolic collision
with the Earth.  Once it has entered the chaotically unstable region
about $L_4$, a collision with the Earth is likely. We will
discuss this phenomenon in detail in Section 4. For references on the
formation of planetesimals and related issues, see
(Goldreich 1973; Goldreich $\&$ Tremaine 1980; Ida $\&$ Makino 1993; Rafikov 2003; 
Wetherill 1989).
We are considering
instability of motion near $L_4$ due to encounters by planetesimals.
(The instability of the Jupiter's outer Trojan asteroids due to the
gravitational effects of Jupiter over time studied 
in Levinson, Shoemaker E. M.  $\&$ Shoemaker C. S. (1997), is a different process.) 

Horseshoe orbits connected with the Earth exist. In fact, an asteroid with
a 0.1 km diameter, 2002 $AA_{29}$, has recently been discovered in just such a
horseshoe-type orbit which currently approaches the Earth to within a
distance of only 3.6 million km (Conners et al. 2002). Horseshoe orbits
about the Sun of this type are also called Earth co-orbiting trajectories,
which are in 1:1 mean motion resonance. An interesting pair of objects in horseshoe orbits about Saturn
are discussed in the Note after the Discussion section. A theoretical 
study of the distribution of objects in co-orbital motion is given by Morais $\&$ Morbidelli (2002).

In this paper we describe a special set of collision
orbits with the Earth which exist due to escape from $L_4$ due 
to planetesimal perturbations. The perturbations cause a gradual
peculiar velocity increase of the mass forming at $L_4$ so that it eventually
achieves a critical escape velocity to send it toward a parabolic Earth
collision approximately in the plane of motion of the Earth about the Sun.
The region in velocity space where escape from $L_4$ occurs in
this fashion is relatively narrow. This mechanism therefore involves
a special set of $L_4$
ejection trajectories which creep towards collision with the Earth.
At the end of Section 3 we will present a full simulation in
three-dimensions of a collision of a Mars-sized impactor with the Earth
assuming a thin planetesimal disk, using the general three-body
problem, where planetesimal encounters with both the impactor and
the Earth are done in a random fashion. The Appendix of this paper
discusses the dynamics of the random planetesimal encounters.

The paper has several main results: 

\noindent We show that a stable orbit at $L_4$ exists where a Mars-sized giant impactor could
grow by accretion without colliding with the Earth. We show that eventually perturbations by other
planetesimals can cause the giant impactor to escape from $L_4$ and send it onto a horseshoe orbit and then
onto a creeping chaotic trajectory with an appreciable probability of having a near parabolic
collision with the Earth. In the Discussion section we argue how this scenario fits in extremely well 
with giant impactor theory and explains the identical oxygen isotope abundances of the Earth and the Moon.
The solar system itself provides a testing ground for our model. As we have mentioned the Trojan asteroids
show that planetesimals can remain trapped at Lagrange points, and in the "Note added" we point out that
the system of Saturn's moons provide examples where the phenomenon we are discussing can be observed,
supporting our model. Finally both in the "Note added" and in the Appendix we discuss prospects for future
work.   

The spirit of this paper is to suggest the intriguing possibility that the hypothesized Mars-sized 
impactor could have originated at $L_4$(or$L_5$). It is hoped that this lays the ground work for more detailed
simulations and work in the future.

\section{Models and a Stability Theorem}
\label{Section:2}

Let $P_1$ represent the Sun, $P_2$ the Earth, and $P_3$ a third mass
particle. We will model the motion of $P_3$ with systems of differential
equations for the restricted and general three-body problems.

\begin{figure}[ht!]
\begin{center}
\resizebox{90mm}{!}
  {\includegraphics{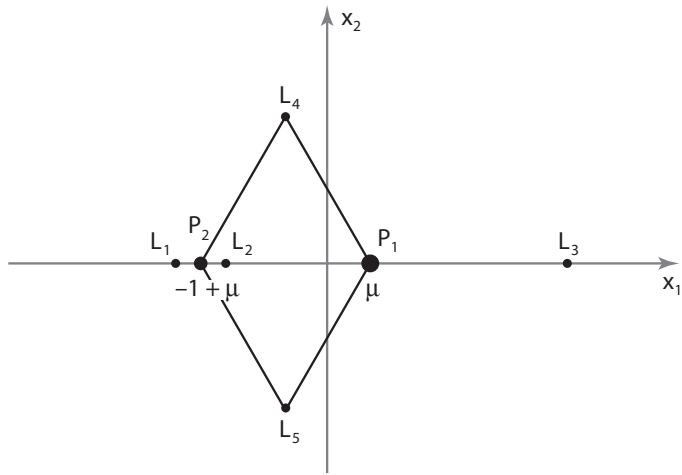}}
\end{center}
\caption{Rotating coordinate system and locations of the
Lagrange points.}
\label{Figure:figureA}
\end{figure}

The first preliminary model, and key for this paper, is the planar circular restricted
three-body problem which assumes the following: 1. $P_1, P_2$ move in
mutual Keplerian circular orbits about their common center of mass which
is placed at the origin of an inertial coordinate system $X,Y$. 2. The
mass of $P_3$ is zero. Thus, $P_3$ is gravitationally perturbed by  $P_1,
P_2$, but not conversely. Letting $m_k$ represent the masses of $P_k$,
$k=1,2,3$, then $m_3 = 0$, and we assume that $m_2/m_1 = .000003$.
Let $\omega$ be the constant frequency of circular motion of $m_1$ and
$m_2$, $\omega = 2\pi/P$, where $P$ is the period of the motion. 
We consider a rotating coordinate system $(x,y)$ which rotates with the constant frequency
$\omega$
as $P_1$ and $P_2$. In the $x$--$y$
coordinate system the positions of $P_1$ and $P_2$ are fixed.  Without
loss of generality, we can set $\omega=1$ and place $P_1$ at $(\mu,0)$
and $P_2$ at $(-1+\mu,0)$.  Here we normalize the mass of $m_1$ to $1-\mu$
and $m_2$ to $\mu$, $\mu =  m_2/(m_1+m_2) = .000003$ .  The equations
of motion for $P_3$ are
\begin{equation}
\begin{array}{lll}
\ddot{x}-2\dot{y} & = & x+\Omega_x \\
\rowspace \ddot{y}+2\dot{x} & = & y+\Omega_y , 
\end{array}
\label{eq:2}
\end{equation}
where $\dot{}\equiv \frac{d}{dt}, \Omega_x \equiv \frac{\partial{\Omega}}{\partial
x}$, 
\begin{displaymath}
\Omega=\frac{1-\mu}{r_1}+\frac{\mu}{r_2} ,
\end{displaymath}
$r_1=$ distance of $P_3$ to $P_1=[(x-\mu)^2+y^2]^{\frac{1}{2}}$, and
$r_2=$ distance of $P_3$ to $P_2=[(x+1-\mu)^2+y^2]^{\frac{1}{2}}$,
see Figure \ref{Figure:figureA}. The right hand side of (\ref{eq:2})
represents the sum of the radially directed centrifugal force
${\bf F_C}=(x,y)$ and the sum ${\bf F_G} = (\Omega_x,\Omega_y)$ of
the gravitational forces due to $P_1$ and $P_2$. 
We note that the units of position, velocity and time
are dimensionless. To obtain position in kilometers, the dimensionless
position $(x,y)$ is multiplied by $149,600,000$ which is the distance
of the Earth to the Sun. To obtain the velocity in km/s, s = seconds,
the velocity $\dot{x},\dot{y}$ is multiplied by the circular velocity
of the Earth about the Sun, $29.78$ km/s.  For (\ref{eq:2}), $t=2\pi$
corresponds to 1 year.

It is noted that (\ref{eq:2}) is invariant under the transformation, $x
\rightarrow x, \ \ y \rightarrow -y, \ \ t \rightarrow -t$. This implies
that solutions in the upper half-plane are symmetric to solutions in the
lower-half plane with the direction of motion reversed. This implies, as
noted in the introduction, that all the results we will obtain for $L_4$
are automatically true for $L_5$, and thus $L_4$ need only be considered.

System (\ref{eq:2}) of differential equations has five equilibrium points
at the well known Lagrange points $L_k, k=1,2,3,4,5$, where $\ddot{x}=\ddot{y}=0$
and $\dot{x}=\dot{y}=0$. (This implies that ${\bf F_C} {\bf +} {\bf F_G}
= {\bf 0}$.)  Placing $P_3$ at any of these locations implies it will
remain fixed at these positions for all time. The relative positions
of $L_k$ are shown in Figure \ref{Figure:figureA}. The locations of
the Lagrange points for arbitrary $\mu \in (0,1)$ are a function of
$\mu$. Three of these points are collinear and lie on the $x$-axis,
and the two that lie off of the $x$-axis are called equilateral points.
We note that
the labeling of the locations of the Lagrange points varies throughout the
literature. We are using the labeling consistent with that in
Szebehely (1967), where in Figure \ref{Figure:figureA} $L_2$ is interior to
to $m_2$ and $m_1$, and where $L_4$ lies above the x-axis. (Note: This means 
$L_4$ is $60^o$ behind the Earth.)

The three collinear Lagrange points $L_k, k=1,2,3$ lying on the $x$-axis
are unstable. This implies that a gravitational perturbation of $P_3$ at any
of the collinear Lagrange points will cause $P_3$ to move away from these points as
time progresses since their solutions near any of these points are
dominated by exponential terms with positive real eigenvalues (Conley 1969). The two equilateral
Lagrange points are stable, so that if $P_3$ were place at these points and gravitationally
perturbed a small amount, it will remain in motion near these points for all time. 
This stability result for $L_4, L_5$ is subtle and was a motivation for the development 
of the so called  Kolmogorov-Arnold-Moser(KAM) theorem on the stability of motion of
quasi-periodic motion in 
general Hamiltonian systems of differential equations (Arnold 1961; 1989; Siegel $\&$ Moser 1971).  
A variation of this theorem was applied to the stability problem of $L_4, L_5$ by Deprit $\&$
Deprit-Bartolom\'e (1967). Their result is summarized in the following
result and represents a major application of KAM theory,
\medskip

\noindent
$L_4,L_5$ are locally stable if $0 < \mu < \mu_1, \ \  \mu_1
= {1\over2}(1-{1\over9}\sqrt{69}) \approx .0385,$ and $\mu \neq \mu_k,
k=2,3,4\ \ \mu_2 = {1\over2}(1-{1\over45}\sqrt{1833}) \approx .0243, \
\mu_3 = {1\over2}(1-{1\over15}\sqrt{213}) \approx .0135, \ \mu_4 \approx
.0109.$  

\noindent For further details connected with this result, see (Belbruno $\&$ Gott 2004)

\noindent
In our case, the Earth has $\mu = .000003$ which is substantially 
less than $\mu_1$
and the exceptional values $\mu_k , k=2,3,4$ so that $L_4$ is clearly
stable for the case of the Earth, Sun system.

An integral of motion for (\ref{eq:2}) is the Jacobi energy given by 
\begin{equation}
J = -(\dot{x}^2+\dot{y}^2)+(x^2+y^2)+\mu(1-\mu)+2\Omega .
\label{eq:3}
\end{equation}
Thus $ \Sigma(C) =\{ (x,y,\dot{x},\dot{y}) \in {\R }^4 \, | \,  
J =C, \, \, C\in \R \}$ is a three-dimensional surface in the four-dimensional phase space
$(x, y,\dot{x},\dot{y})$, such that the solutions of (\ref{eq:2})
which start on $\Sigma(C)$ remain on it for all time. $C$ is called the
Jacobi constant. The manifold $\Sigma(C)$ exists in the four-dimensional phase space. It's
topology changes as a function of the energy value $C$. This can be
seen if we project $\Sigma$ into the two-dimensional position space
$(x,y)$.  This yields the Hill's regions $\mathcal{H}(C)$ where
$P_3$ is constrained to move. 
The qualitative appearance of the Hill regions 
$\mathcal{H}(C)$  for different values of $C$ are 
described in Belbruno (2004) and Szebehely (1967). As $C$ decreases in value,
$P_3$ has a higher velocity magnitude at a given point in the $(x,y)$-plane.  

In this paper we will be considering cases where $C$ is slightly less than
3, $C \stackrel{<}{\sim} 3$, where the Hill's region is then the entire plane. Thus, in this case
$P_3$ is free to move throughout the entire plane.  

In the next section we will initially use the planar circular restricted
three-body problem to obtain insight into the motion of $P_3$ near $L_4$.
Ultimately we are interested in the general
three-dimensional three-body problem with the mass points $P_k,
k=1,2,3$ of respective masses $m_k$. Unlike the restricted problem, $m_3$
need not be zero, and $P_1,P_2$ are not defined by constant circular Keplerian
motion. Instead, $P_1,P_2$ will be given {\em initial conditions} for
uniform circular Keplerian motion between the Earth, $P_2$, and Sun,
$P_1$, assuming the Earth is 1AU distant from the Sun. However, for $m_3 \neq 0$ this circular motion
will not be constant. For $m_3$ small,
the deviation of the motion of $P_1,P_2$ from the circular motion
will in general be very small. Later in this paper we will be setting
\begin{equation} 
m_3 \ \ = \ \ .1 m_2, 
\label{eq:4} 
\end{equation}
where $m_2$ is the mass of the Earth, and so $\mu = .000003$. Thus,  $m_3$ is a Mars-sized impactor.

The differential equations for the general three-dimensional three-body
problem in inertial coordinates ($X_1,X_2,X_3$) are defined by
the motion of the 3 mass particles $P_k$ of masses $m_k > 0, k=1,2,3$,
moving in three-dimensional space $X_1, X_2, X_3$ under the classical
Newtonian inverse square gravitational force law. We assume the Cartesian
coordinates of the $k$-th particle are given by the real vector ${\bf
X}_k = (X_{k1}, X_{k2}, X_{k3}) \in {\R^3}$. The differential equations
defining the motion of the particles are given by
\begin{equation}
m_k {\bf \ddot{X}}_k = \sum_{j=1 \atop j \neq k} ^3
\frac{G m_j m_k}{r^2_{jk}} \frac{{\bf X}_j - {\bf X}_k}{r_{jk}}, 
\label{eq:5}
\end{equation}
$k=1,2,3$, where $r_{jk} = |{\bf X}_j - {\bf X}_k| = \sqrt{\sum_{i=1}^3
(X_{ji} - X_{ki})^2}$ is the Euclidean distance between the $k$-th and
$j$-th particles, $G$ is the universal gravitational constant, and $^{.}
\equiv \frac{d}{dt}$. Equation (\ref{eq:5}) expresses the fact that the
acceleration of the $k$-th particle $P_k$ is due to the sum of the forces
of the $2$ particles $P_i, i=1,2, 3,  i \neq k$. The time variable $t
\in {\R^1}$. Without loss of generality, we place the center of mass of
the three particles at the origin of the coordinate system.

We note that the stability result by Deprit $\&$ Deprit-Bartolom\'e (1967) provides conditions for stability of $P_3$ with
respect to $L_4$; however, it does not necessarily provide conclusions
on instability. It is proven more generally in Siegel $\&$
Moser (1971) that in the three-body problem, if $m_3$ satisfies $$
     27(m_1m_2 + m_2m_3 + m_3m_1) > (m_1 + m_2 + m_3)^2
$$ then the motion is unstable. This is not satisfied in our case since
$m_1 = 1 - \mu, \  m_2 = \mu, \ m_3 = .1\mu, \ \mu=.000003$.  However,
if it is not satisfied it does not guarantee stability, and a deeper
analysis is required such as KAM theory. In this more general case a
result like that of Deprit $\&$ Deprit-Bartolom\'e (1967) is not available.

We have verified in the general three-dimensional three-body problem defined 
by (\ref{eq:5}) with $m_1 = 1 - \mu, \  m_2 = \mu, \ m_3 = .1\mu, \ \mu=.000003$,
and more generally in the three-dimensional model for the solar system that
$L_4$ is stable for a numerical integration time span of 10 million years. The model of the solar
system we used includes the nine planets and is modeled as an n-body 
problem with
circular coplanar initial conditions using the current masses of the planets and
radii of the planetary orbits. The integration time span of 10 million years is suitable for 
the purposes of our analysis.

\section{Chaotic Creeping Orbits Leading to Parabolic Earth Collision}
\label{Section:3} 

While we have shown via full solar system that $L_4$ is stable it might be
argued that this stability could be perturbed by other planetesimals, and in 
fact it is exactly this process that we are investigating (see Appendix). We 
expect that gravitational perturbations from other planetesimals will, via a 
random walk process in peculiar velocity, cause the Mars-sized impactor 
to eventually escape from $L_4$. 
We numerically demonstrate in this section that there exists a family
of trajectories leading from $L_4$ to parabolic Earth collision.
Producing these trajectories shows that Earth
collision is likely when $P_3$ escapes $L_4$. $P_3$ escapes
$L_4$ once it achieves a critical peculiar velocity - in the rotating
frame.

To describe the construction of the parabolic Earth colliding
trajectories, we will begin first with the planar restricted
problem. Then, we will show that the results hold up as we make the
model more realistic.

\medskip

Assuming $m_3 = 0$ we consider System (\ref{eq:2}) and place $P_3$
precisely at $L_4$.  As long as the velocity of $P_3$ relative to $L_4$
is zero, then $P_3$ will remain at $L_4$ for all time.

The velocity vector at $L_4$ for $P_3$ is given by  ${\bf{v}} = (\dot{x},
\dot{y})$. Let $\alpha \in [0, 2\pi]$ be the angle that $\bf{v}$ makes
with the local axis through $L_4$ that is parallel to the $x$-axis.  Thus,
${\bf{v}} = V * (\cos{\alpha}, \sin{\alpha}), \ \ V \equiv |{\bf v}|$ $=
\sqrt{{\dot{x}}^2 + {\dot{y}}^2}$.

When $V \neq 0$ and if $t=0$ is the initial time for $P_3$ at $L_4$,
then for $t > 0$, $P_3$ need not remain stationary at $L_4$.  If $V$
is sufficiently small, then by the result of Deprit $\&$ Deprit-Bartolom\'e (1967) the velocity of $P_3$
should remain small for all $t > 0$, and $P_3$ should remain within a
small bounded neighborhood of $L_4$. This follows by continuity with
respect to initial conditions. However, as $V(0) \equiv V(t)|_{t=0}$
increases, then the resulting motion of $P_3$ need not stay close to $L_4$
for $t > 0$. This is investigated next.

\medskip

We fix $\alpha$ and fixing $P_3$ at $L_4$ at $t=0$, we gradually increase
$V(0)$ and observe the motion of the solution curve $\gamma(t) = (x(t),
y(t))$ for $t > 0$ for each choice of $V(0)$.  This is done by numerical
integration of System (\ref{eq:2}). (All the numerical integrations in
this paper are done using the numerical integrator NDSolve
of Mathematica 4.2 until further notice.)  The following general results are obtained which we
first state, and then illustrate with a number of plots
(In all of the plots of orbits of the restricted problem (\ref{eq:2}) in the $x,y$
plane which are labeled 'Sun centered' the translation $x \rightarrow x + \mu,
y \rightarrow y$ has been applied which puts the Sun at the origin, and the Earth at
the point $(-1,0)$). 

\medskip
\noindent
{\bf R1}
\noindent 
For each choice of $\alpha \in [0, 2\pi]$, as $V(0)$ is gradually
increased from $V(0) = 0$, and where $\gamma(0) = (x(0), y(0))$ is at
$L_4$, the trajectory $\gamma(t)$ for $t > 0$, remains in small arc-like regions about
$L_4$, which as $V(0)$ increases, evolve into thin horseshoe regions
containing $L_4$ and lying very near to the Earth's orbit about the Sun. As
$V(0)$ increases further, the horseshoe region begins to close on
itself,
approaching forming a continuous annular ring about the Sun, coming close to
connecting at the Earth. It is found that there exists a well defined
critical value of $V(0) = V^*(0)$ where the ring closes at the Earth,
and then the motion of $P_3$ bifurcates from a motion constrained to the
horseshoe-like region where it never makes a full cycle about the Sun, to
a motion where it continuously cycles
about the Sun, repeatably passing close to the Earth, and no longer in
the horseshoe-like motion.  $V^*(0)$ has an approximate value for most values of
$\alpha$ between .200 km/s to .600 km/s. 
We refer to this continuously cycling motion for $V(0) = V^*(0)$ as
{\em breakout}. Breakout continues to occur for $V(0) > V^*(0)$ \medskip

\medskip
\noindent
{\bf R2}
\noindent
In breakout motion for $V(0) \stackrel{>}{\sim} V^*(0)$ or $V(0)=V^*(0)$, the trajectory $\gamma(t)$
for $t > 0$ traces out a dense set of orbits in a thin annular region
repeatably passing near the Earth, where the fly-bys at Earth periapsis
appear to be all approximately parabolic. 
The breakout orbit is chaotic in
nature so that small changes in $V(0)$ result in breakout trajectories
which are in general significantly different in appearance, and still restricted
to a thin annular region about $P_1$. The
breakout trajectories as they cycle about the Sun have a high
likelihood of colliding with the Earth. Moreover, for each $\alpha$ a near
parabolic collision trajectory is readily found for $V(0) \stackrel{>}{\sim}
V^*(0)$. The collision orbits move near the Earth's orbit and gradually
approach the Earth for collision. (This gradual motion approximately
along the Earth's orbit, we refer to as {\em creeping}.) The collision
orbits can creep to collision along the direct or retrograde directions
with respect to $P_1$.

\medskip
\noindent
{\em Demonstration of R1}
\medskip

\begin{figure}[ht!]
\begin{center}
\resizebox{100mm}{!}
  {\includegraphics{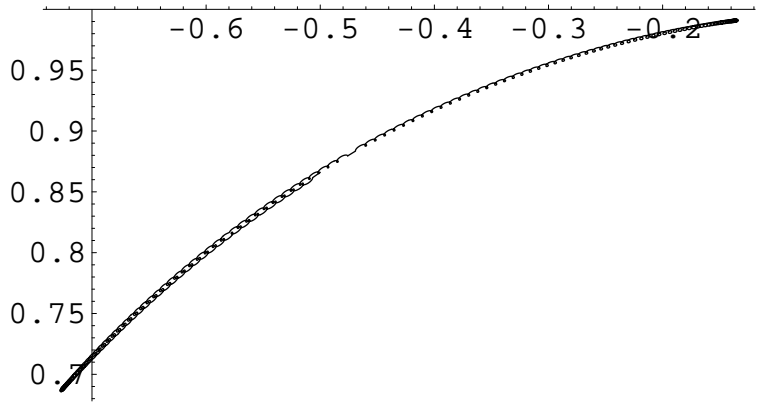}}
\end{center}
\caption{$\gamma(t), V(0) = .001, t\in [0,1000]$, x vs y (i.e. x-axis is
horizontal, y-axis is vertical), Sun centered.}
\label{Figure:V001}
\end{figure}

We choose an arbitrary velocity direction
${\bf v}$ for $P_3$ at $L_4$ at $t=0$, where ${\bf v}$ points in the
vertical positive $y$ direction. In our simulations, the location of $L_4$ is
$(-.5,{\sqrt{3}\over2})$,  and the $y-$coordinate is input with the value .866025404.
Beginning with $\alpha = \pi/2$, we choose
a magnitude $V = V(0) = .001$, and numerically integrate the system
of differential equations (\ref{eq:2}) forward for $t \in [0,1000]$.
This velocity magnitude is small, and since $L_4$ is stable $P_3$
remains in a thin arc-like region approximately of radius 1 shown in
Figure \ref{Figure:V001}. $P_3$ starts at the location 
\noindent
${\bf x}(0) = (.5, .5\sqrt{3})$, and moves down in the posigrade direction with respect
to the Sun. As it moves, it performs many small loops as are shown
in Figure \ref{Figure:V001}. These loops occur since the semi-major axis
of the orbit of $P_3$ has changed slightly from 1 and the orbit of $P_3$
has a slight nonzero ellipticity, both due to the addition of $V(0)$.
So, as it moves in its approximate elliptical motion over the course
of one year it falls slightly behind and forward with respect to the
Earth when it is at its apoapsis and periapsis, respectively. Each loop
forms in one year. Thus, for $t=1000$, there are $1000/(2\pi)$ loops.
$P_3$ moves down to a minimal location where $y$ is approximately .7,
and then it turns around and moves in the upward direction where the
small loops point in the opposite direction when it was moving in the
downward direction.  The superposition of the loops makes a braided
pattern as seen in the lower half of Figure \ref{Figure:V001}. $P_3$
stays in this bounded arc-like region since $L_4$ is stable, and the
velocity $V(0)$ is relatively small. (If $V(0)=0$, then $P_3$ stays fixed
at $L_4$ for all time.) Because the velocity magnitude is small, $P_4$
has a Kepler energy nearly that of $L_4$, and so its semi-major axis
with respect to the Sun deviates from 1 by a negligible amount.  Thus,
as it moves, it stays nearly on a circle of radius 1. That is, in an
inertial coordinate system it stays approximately on Earth's orbit about
the Sun.  As long as $V(0)$ is small, which it is throughout this paper,
the trajectories of $P_3$ remain close to the Earth's orbit and move
with small loops, in the rotating coordinate system. The particle $P_3$
creeps slowly along the Earth's orbit initially in a posigrade fashion,
and then in a retrograde fashion away from the Earth.

\begin{figure}[ht!]
\begin{center}
\resizebox{70mm}{!}
  {\includegraphics{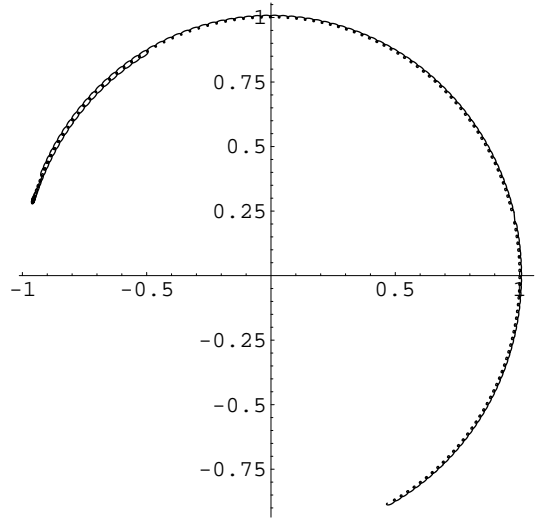}}
\end{center}
\caption{$\gamma(t), V(0) = .004, t\in [0,1000]$, x vs y, Sun centered.}
\label{Figure:V004}
\end{figure}

\begin{figure}[ht!]
\begin{center}
\resizebox{70mm}{!}
  {\includegraphics{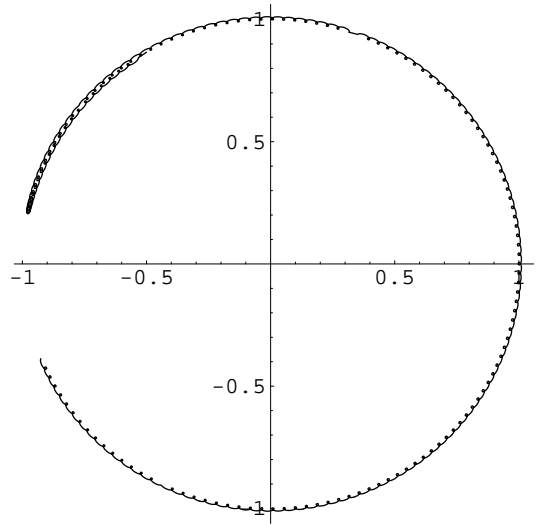}}
\end{center}
\caption{$\gamma(t), V(0) = .005, t\in [0,1000]$, x vs y, Sun centered.}
\label{Figure:V005}
\end{figure}

The above procedure is repeated, where we slightly increase the value of
$V(0)$ to .004 at $L_4$ at $t=0$. Since $V(0)$ has increased, then as is
seen in Figure \ref{Figure:V004}, where $t \in [0,1000]$, $P_3$ creeps
\noindent
further in its Earth-like orbit about the Sun. Since $V(0)$ is small,
the trajectory of $P_3$ deviates slightly from a circle of radius 1. This
deviation slightly increases as $V(0)$ increases. The addition of $V(0)$
at $L_4$ causes $P_3$ to have a slightly smaller value of the Jacobi
integral, to be slightly less than 3 ($C\stackrel{<}{\sim}3$). This
means
that $P_3$ becomes more energetic, and thus can creep further along the
Earth's orbit.  Increasing $V(0)$ by .001 to .005 causes the increased
creeping shown in Figure \ref{Figure:V005}, where $t \in [0,1000]$.
\begin{figure}[ht!]
\begin{center}
\resizebox{70mm}{!}
  {\includegraphics{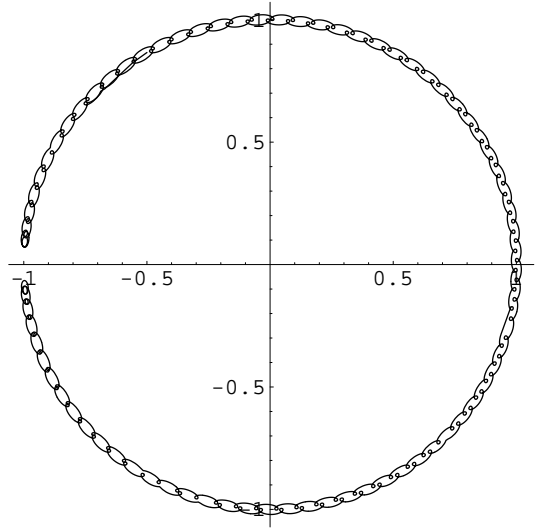}}
\end{center}
\caption{$\gamma(t), V(0) = .009, t\in [0,1000]$, x vs y, Sun centered.}
\label{Figure:V009}
\end{figure}

In Figure \ref{Figure:V009}, $V(0)$ is increased to .009. $P_3$ leaves $L_4$, moves downward in a
posigrade fashion to slightly behind the Earth, then turns around and moves in a retrograde fashion 
on its Earth-like orbit about the Sun, until it approaches the Earth from the front turning around
and then moving in a posigrade fashion. 

A braided pattern results due to the fact the Earth-like orbit 
is traversed twice, with loops pointing in the inner and outer directions. The resulting complicated
looking trajectory is symmetric with respect to the $x$-axis due the symmetry mentioned earlier for the
restricted problem. The width of the region near the Earth's orbit in which $P_3$ moves has slightly
increased due to the increase in $V(0)$.

We note that general appearance of the orbit of the asteroid 2002 $AA_{29}$, mentioned in the
introduction (Conners et al. 2002), remarkably is very similar in appearance
to Figure \ref{Figure:V009}. Unlike the planar orbit considered here, 2002 $AA_{29}$ has a inclination of
10 degrees with respect to the plane of the Earth's orbit. It's 
oscillation period is approximately 95 years. The approximate period of the orbit in 
Figure \ref{Figure:V009} is about 159 years, which is not too dissimilar.  
Orbits of this type are called {\em horseshoe orbits}. The horseshoe orbits are constrained to a region we
refer to as a horseshoe region so $P_3$ cannot move past the Earth. Such a region is discussed in 
Murray $\&$ Dermott (1999). Let $\theta$
be the polar angle measured from the positive $x$-axis for the position of $P_3$. The horseshoe orbits 
have the the property that $\theta \neq \pi$. This means that $P_3$ will not fly by the Earth.
For other papers on this motion, see
(Christou 2000; Hollabaugh $\&$ Everhart 1973; Mikkola $\&$ Innanen 1990; Namouni 1999;
Weissman $\&$ Wetherill 1974).

\begin{figure}[ht!]\begin{center}
\resizebox{70mm}{!}  {\includegraphics{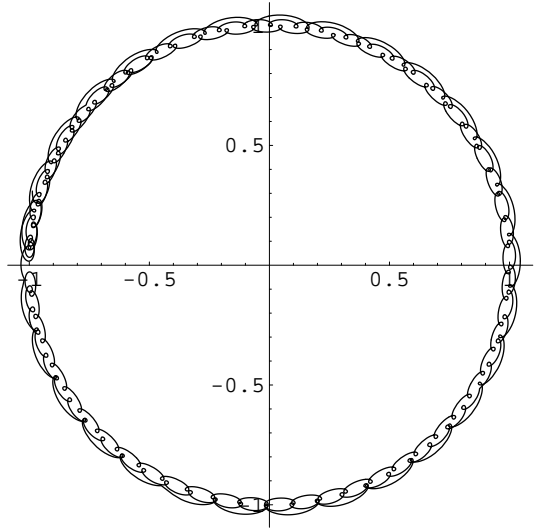}}
\end{center}\caption{$\gamma(t), V(0) = .011, t\in [0,1000]$, x vs y,
Sun centered.}
\label{Figure:V011}
\end{figure}

When $V(0)$ reaches $.011$, $P_3$ is able to escape from the thin horseshoe-like region
and fly by the Earth as is seen in Figure \ref{Figure:V011}. 

\noindent
This achieves breakout motion
where $P_3$ then cycles about the Sun only in one direction. In  Figure \ref{Figure:V011} the
cycling is in the retrograde direction. This actual cycling is not shown in this figure since
for the time range given, breakout into cycling motion occurs when $t=988$, on the outer
retrograde trajectory. $P_3$ first leaves $L_4$ moves near the Earth, then back up
in a retrograde fashion going all the way around the Sun to near and in front of the Earth, then
moving around the Sun again in a posigrade fashion to its location behind the Earth,
then finally it moves on the outer trajectory in a retrograde fashion back to just ahead of the Earth 
when it crosses by the Earth at  $t=988$ (crossing the $x$-axis near the Earth), then performing 
the cycling breakout motion after that time.  
This transition from creeping horseshoe motion to creeping breakout motion is what is desired
for this paper. The transition from horseshoe motion to breakout motion represents
a bifurcation from one type of motion to a different type.  
We are interested in the likelihood of Earth collision while in breakout motion just
after the bifurcation. This represents breakout motion with minimal energy.

This transitional breakout motion has two important properties:  \\
\noindent
1. $P_3$ moves in a thin annular region about the Sun, \\ 
\noindent
2. $P_3$ repeatably flys by the Earth.\\

\noindent These properties imply the following:
Since the annular region is thin, the Earth fly-bys are in general
close. The close Earth fly-bys are approximately parabolic in
nature, as we will demonstrate, and as $P_3$ flys by the Earth its
actual velocity vector is approximately tangent with the Earth's orbit. This
implies that $P_3$ gains a negligible velocity increase due to gravity
assist as it flys by the Earth as we will show.  This guarantees that
$P_3$ will continue to move in an Earth-like orbit about
the Sun, and continue to cycle.  This implies that as $P_3$ moves around
the Sun, it will densely fill the thin annular region it moves in. This
means that it has a high likelihood of colliding with the Earth. We will
demonstrate that collision readily occurs in these creeping breakout
orbits.

In fact, the previous case where $V(0)=.011$, which is the first breakout
motion we computed, leads immediately to collision at $t = 1384.7176$
(or 220.3847 years).
In our exposition below, we will use a slightly different value of $V(0)$
which happens to achieve collision at an even earlier time.

\begin{figure}[ht!]
\begin{center}
\resizebox{70mm}{!}
  {\includegraphics{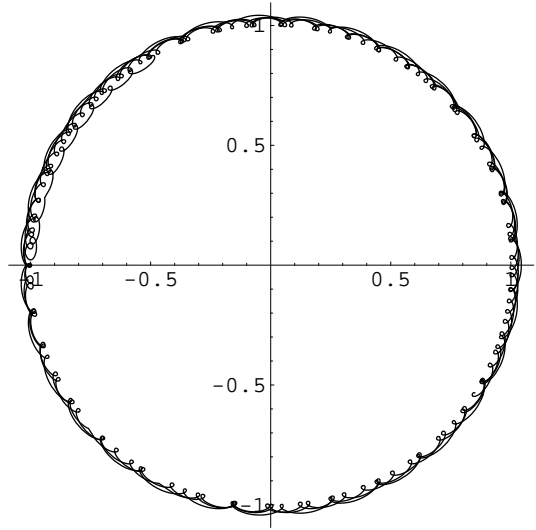}}
\end{center}
\caption{$\gamma(t), V(0) = .012, t\in [0,1000]$, x vs y, Sun centered.}
\label{Figure:V012}
\end{figure}

Breakout motion is seen in Figure \ref{Figure:V012}. It is observed that
a shift
from $V(0)=.009$ to $V(0)=.012$ causes a qualitatively different looking picture, where the
bifurcation between horseshoe and breakout motion is clearly seen.

\begin{figure}[ht!]
\begin{center}
\resizebox{70mm}{!}
  {\includegraphics{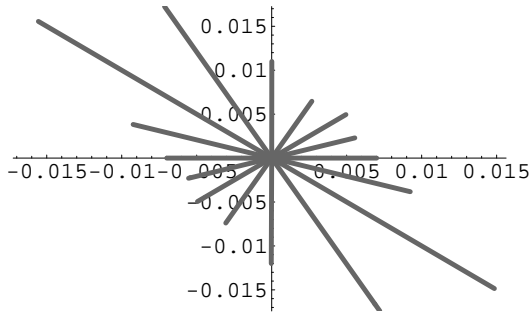}}
\end{center}
\caption{Initial velocity directions ${\bf v}(0)$ at $L_4$
whose magnitude corresponds to the
associated critical breakout velocity $V^*(0)$, which are listed in
Table
\ref{tab:1}.}
\label{Figure:ScatteringDirections}
\end{figure}

The case just considered is for the direction $\alpha=\pi/2$. The
same procedure produces critical values of $V(0)=V(0)^*$ leading to
breakout motion, from horseshoe motion, for any value of $\alpha \in
[0,2\pi]$. A set of these for $\alpha$ increments of $\pi/8$ are
listed in (Belbruno $\&$ Gott 2004) in Table 1. This is graphically shown in Figure
\ref{Figure:ScatteringDirections}.  In this figure, the length of
each line is equal to the value of $V^*(0)$ in that direction. In this way, a
smooth variation of $V^*(0)$ as a function of $\alpha$ is numerically
obtained. There is a sharp spike in the value of $V^*(0)$ which has
a maximum at $5.102\pi/8$ of .22. There is also a similar maximum
near the value of $13.5\pi/8$. These are not listed since they are
not typical: almost all the values of $V^*(0)$ are in the range of
values illustrated. The minimum value of $V^*(0) = .0057$ is for $\alpha =
(9\pi/8)-.01$. Multiplying the values of $V^*(0)$ by 29.78 yields a
range of velocity values generally between 180 meters/s and 1.2 km/s.
Note that the two directions corresponding to the maximal spikes in velocity
seen in Figure \ref{Figure:ScatteringDirections} approximately lie near
the Sun and anti-Sun directions. In this figure the Sun is toward the
lower right.

Note that the method, or algorithm, used above to estimate
the critical velocities $V^*(0)$ at $L_4$ leading to breakout motion,
is similar in nature to the method of estimating transitional stability
regions, called weak stability boundaries, between capture and escape about the Moon described in
Belbruno (2004). This capture region has important applications.
It was used by one of us to find a new type of low energy route
to the Moon in 1990 where lunar capture is automatic (Belbruno $\&$ Miller 1990).
This special lunar transfer was designed in order to resurrect a
Japanese lunar mission and enable the spacecraft {\it Hiten} to successfully
reach the Moon in October 1991 with almost no fuel
(Belbruno 1992; Frank 1994). More general references on
this are (Adler 2000; Belbruno 2004; Belbruno $\&$ Miller 1993). 

Other methods could be used to study the bifurcation from horseshoe to
breakout motion such as the computation of suitable surfaces of section to the
trajectories in phase space, and then monitoring the iterates of
intersecting trajectories on the section. This would give a more complete knowledge
of the phase space near breakout motion, but this approach is not necessary for
our purposes. The algorithm we have described  accurately determines when 
bifurcation occurs.

We have also performed an analysis to understand the relationship of the critical
breakout velocities as a function of the mass of the Earth $m_2$. It is
found that roughly
$$V^*(0) \propto m_2^{1/3}.$$
As an example, we consider the two cases: $m_2=.1m_E,
.01m_E, \ \ m_E = .000003$. When $m_2=.1m_E$, then for $\alpha =
0,\pi/2,\pi,3\pi/2$, we obtain $V^*(0)=.004,.007,.004,.007$,
respectively, and for $m_2=.01m_E$, we obtain
$V^*(0)=.002.,003,.002,.003$,
respectively.\ \ These results imply that $V^*(0) \sim 0.6 m_2^{1/3}$ for
$\alpha = 0,\pi$, and when
$\alpha = \pi/2,3\pi/2, \ \ V^*(0) \sim  m_2^{1/3}$. (Thus a giant impactor trapped
in a stable orbit about $L_4$ and unperturbed will remain trapped there as the
proto-Earth grows by accretion. Breakout velocity increases as the
proto-Earth grows, postponing breakout, but at late times after 
the proto-Earth has reached essentially its full mass, according to our
scenario, perturbations can drive the giant impactor 
to breakout.)

This concludes the demonstration of R1.

\medskip

\noindent
{\em Demonstration of R2}
\medskip

\begin{figure}[ht!]
\begin{center}
\resizebox{70mm}{!}
  {\includegraphics{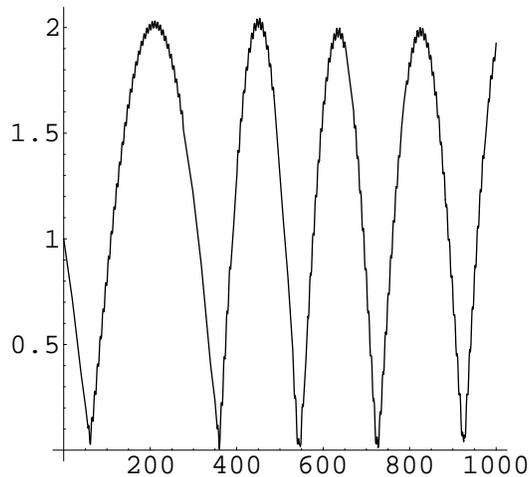}}
\end{center}
\caption{Variation of distance $r_2$ of $P_3$ to $P_2$(earth) as a
function of $t \in [0,1000]$ in dimensionless units,
$V(0) = .0119981$.}
\label{Figure:ExtendedRadius}
\end{figure}

We first show how to readily find trajectories from $L_4$ which collide with the Earth.
The value of $\alpha =\pi/2$ is again considered, and we consider
the case $V(0)=.012 \stackrel{>}{\sim} V^*(0)$ shown in Figure \ref{Figure:V012}.
Plotting the distance $r_2$ between $P_3$ and the Earth for $t \in [0,1000]$ reveals the times of the various
Earth fly-bys. It was found that the case of  $V(0)=.012$, for the given range of
$t$, had very close Earth fly-bys, but no actual collision. Randomly altering this 
value of $V(0)$ yielded a collision on our second random choice
of values of $V(0)=.0119981$. This is seen by plotting $r_2$ as a function of time shown in
Figure \ref{Figure:ExtendedRadius} 

\begin{figure}[ht!]
\begin{center}
\resizebox{70mm}{!}
  {\includegraphics{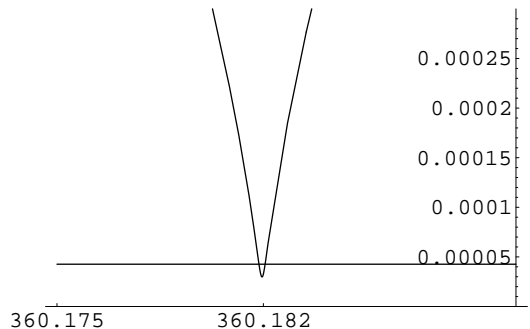}}
\end{center}
\caption{Distance $r_2$ of $P_3$ to $P_2$(earth) goes below the earth's
radius .0000426346(i.e. 6378.14 km) proving collision has occurred.
$V(0)=.0119981$. Plot is $t$ vs $r_2$ in
dimensionless units.}
\label{Figure:CollisionRadius}
\end{figure}

By magnifying the regions near minima of $r_2$ it can be seen which ones
may yield collision.  In this case, counting from the left to the right,
we determined that the first, fourth, fifth minima yield distant fly-bys
at over 1 million km. Thus, in these cases, the Earth fly-bys are not of
interest. The third fly-by misses the Earth by about 18,000 km. However,
it is found that the second fly-by in fact collides with the Earth. 
The time of collision with the surface of the Earth
is at $t = 360.181558$, corresponding to 57.3247 years.
This time is calculated when the center of the impactor, viewed as
a circle of radius $r_I = 3397$ km, intersects the surface boundary of the
Earth, which is at a radial distance $r_E = 6378.14$ km, 
from the Earth's center. This is seen in Figure \ref{Figure:CollisionRadius}.

The horizontal line
indicates the Earth's radius of 6378.14 km here represented by the value
$r_2=.0000426346$. The time of collision with the surface of the Earth
is at $t = 360.181558$, corresponding to 57.3247 years. We are assuming
that each point of the trajectory of the impactor at any given time 
is at the center of the impactor.
More accurately, however, collision actually  
occurs a few moments earlier when the surface boundary, circle, of the impactor 
touches the surface boundary, circle, of the Earth. That is, when
$r_2 = r_I + r_E = 9775.14$ km, where for $r_I$ we take the radius of
Mars, since this is a Mars-sized object, or, in dimensionless units,
when $r_2 = .00065342$. In general, we assume collisions mathematically
occur in our numerical simulations when $r_2 \leq .00065342$.

\begin{figure}[ht!]
\begin{center}
\resizebox{70mm}{!}
  {\includegraphics{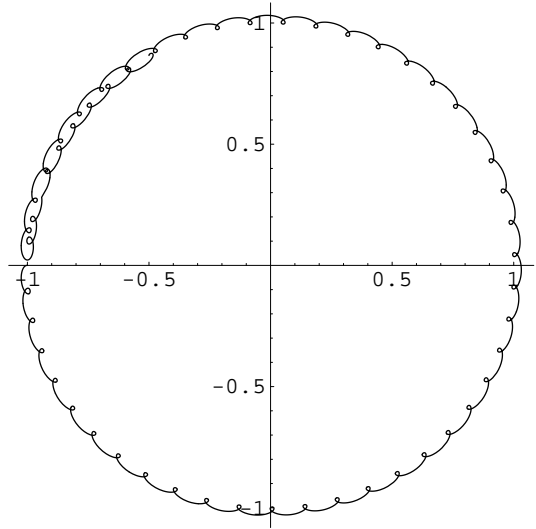}}
\end{center}
\caption{Entire collision orbit (x vs y)- Originating at $L_4$ at $t=0$
and colliding with the earth
when $t = 360.18$ or, equivalently, $57.32$ years,  $V(0)=.0119981$,
Sun centered. }
\label{Figure:EntireCollisionOrbit}
\end{figure}

We now show what the collision trajectory looks like and discuss its properties.
The collision trajectory, $Cl$, is shown in Figure \ref{Figure:EntireCollisionOrbit}.

\begin{figure}[ht!]
\begin{center}
\resizebox{70mm}{!}
  {\includegraphics{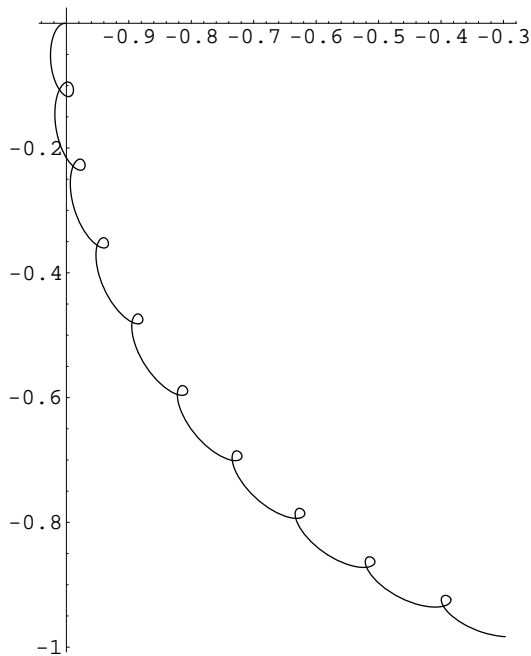}}
\end{center}
\caption{Orbit approaching collision with the earth (x vs y). Time
duration of 9.57 years shown, $t \in [300, 360.181558]$, axis earth
centered.}
\label{Figure:ApproachingCollision}
\end{figure}

It starts at $L_4$, moves in a posigrade fashion toward the Earth, turns around and
in a retrograde motion moves around the Sun to collide with the Earth. A view of this
orbit in its final 9.57 years is shown in Figure \ref{Figure:ApproachingCollision}.

Collision with the Earth itself and the final 9.14 hours of the trajectory are shown in
Figure \ref{Figure:Collision}. 

\begin{figure}[ht!]
\begin{center}
\resizebox{70mm}{!}
  {\includegraphics{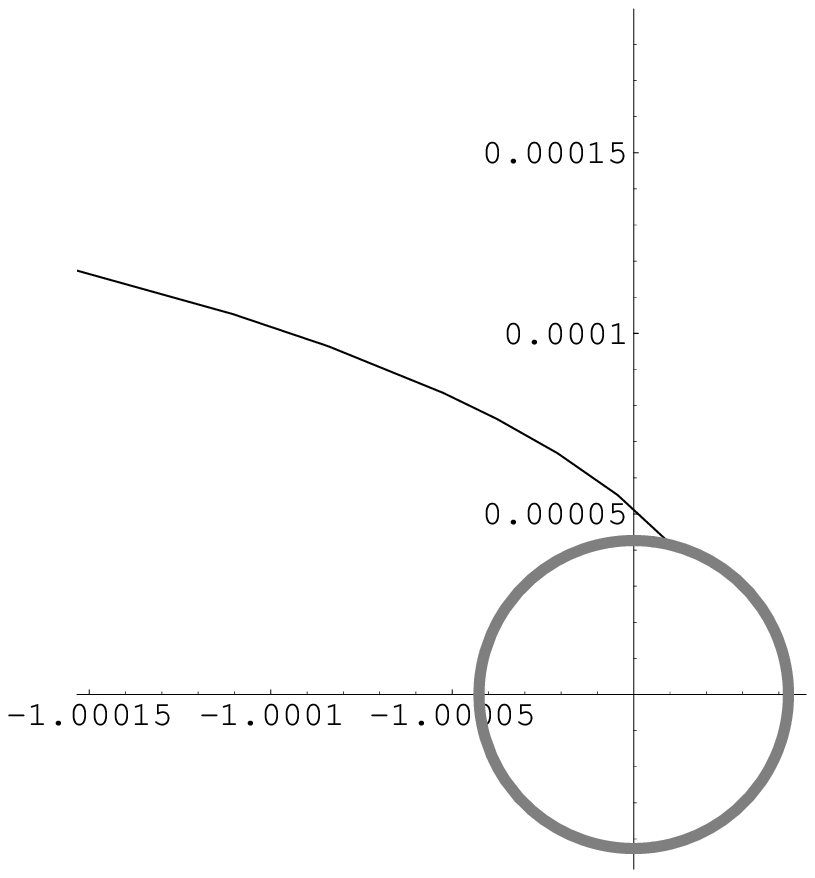}}
\end{center}
\caption{Collision of the impactor with the earth, x vs y.
Final 9.14 hours of trajectory shown, earth
centered.}
\label{Figure:Collision}
\end{figure}

\noindent
In this figure we stopped the trajectory of the impactor before its
Earth periapsis. 
However, if it were continued beyond collision it would reach its periapsis point approximately on the
$x$-axis inside the Earth's radius at $t=360.1817$ where $r_2=.00003$. (See
Figure \ref{Figure:Pass2}). 

\begin{figure}[h!]
\begin{center}
\resizebox{70mm}{!}
  {\includegraphics{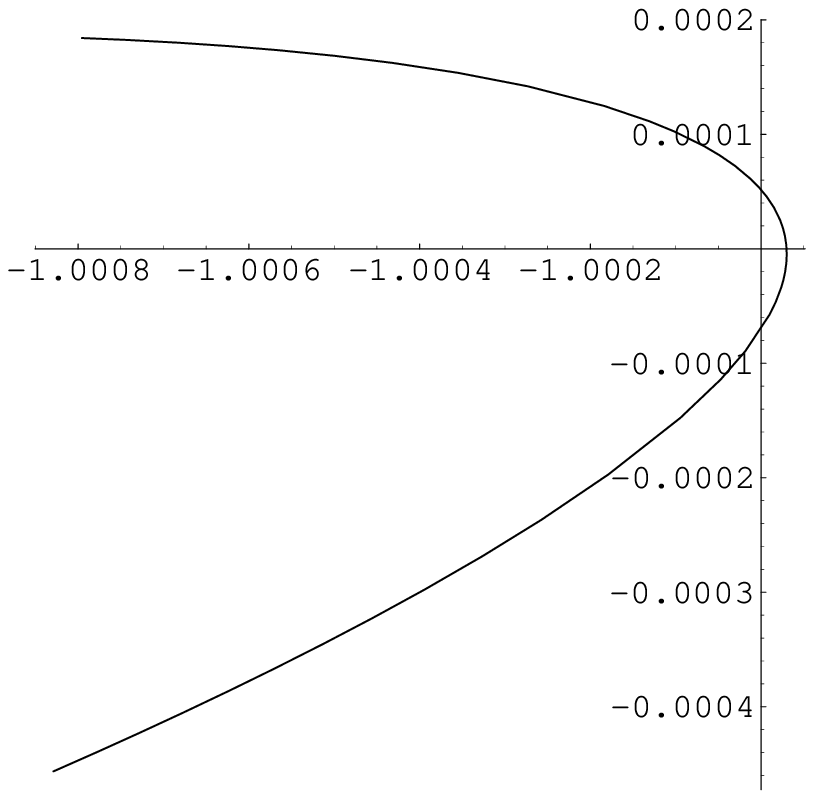}}
\end{center}
\caption{Continuation of collision orbit
across the $x$-axis when $t=360.1816$, x vs y, earth centered.}
\label{Figure:Pass2}
\end{figure}

\begin{figure}[ht!]
\begin{center}
\resizebox{70mm}{!}
  {\includegraphics{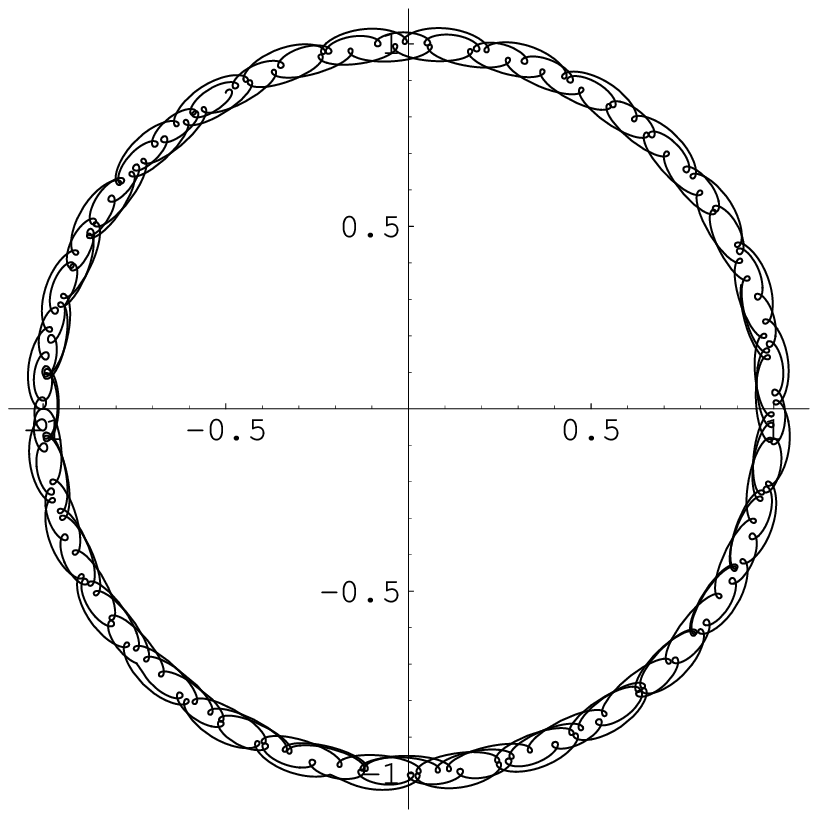}}
\end{center}
\caption{ Extended collision orbit, through collision, $t \in [0,1000]$
( i.e. $t \in [0,159.15]$ years), x vs
y, Sun centered.}
\label{Figure:ExtendedCollision}
\end{figure}

\noindent
If it were extended so that $t \in [0,1000]$,
the trajectory would be as shown in Figure \ref{Figure:ExtendedCollision}.

\begin{figure}[h!]
\begin{center}
\resizebox{70mm}{!}
  {\includegraphics{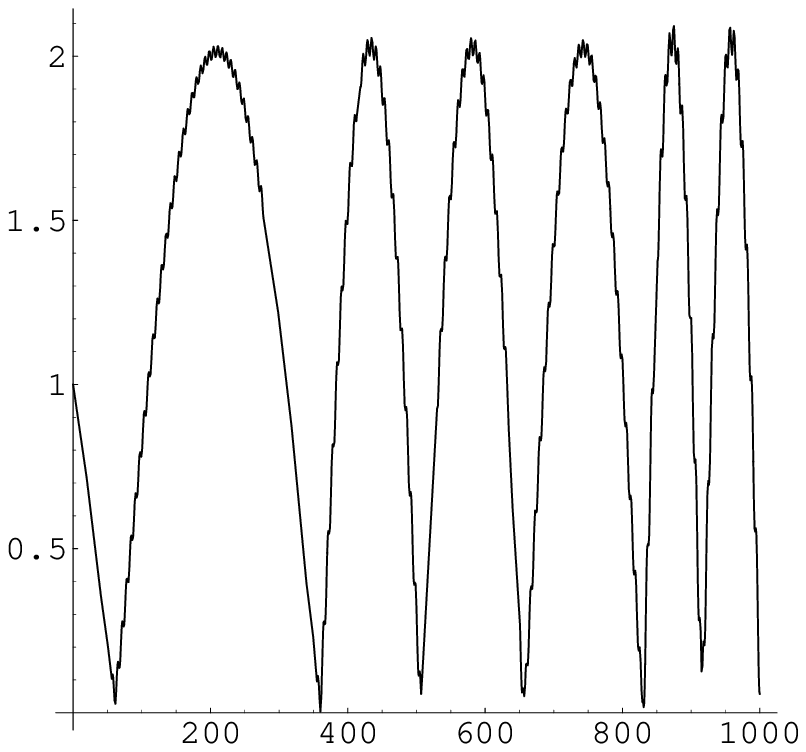}}
\end{center}
\caption{Variation of $r_2, t\in [0,1000]$, for $V(0)=.0119986$. Compare
with Figure \ref{Figure:ExtendedRadius}, where $V(0)$ differs by
.0000005.}
\label{Figure:ExtendedRadius2}
\end{figure}

\noindent
This figure for 
$V(0)=.0119981$ is shown to compare with Figure \ref{Figure:V012} for $V(0)=.012$
indicating the sensitive, or chaotic, nature of the breakout motion, where
a difference of $V(0)$ by .000002 yields a qualitatively different appearing trajectory.
The chaotic nature of the motion near breakout is also seen in 
Figure \ref{Figure:ExtendedRadius2}, which is a plot of $r_2$ for
$V(0)=.0119986$, when compared to Figure \ref{Figure:ExtendedRadius}
where $V(0)=.0119981$. 

It is seen that although the difference in
$V(0)=.0000005$ at $L_4$, there is a significant difference in the qualitative
appearance of the two plots. This is caused by the fact that infinitesimally
small changes in $V(0)$ at $L_4$ can cause slightly different Earth fly-by 
conditions which over long time spans, $t \in [0,1000]$ can cause the trajectory
to change noticeably if any of the fly-bys are close. However, as we will see in the following,
close fly-bys will only yield negligible Kepler energy increases with respect to the Sun.
So, although the trajectory may have a qualitative different appearance,
it will still have approximately the same Kepler energy before and after close fly-bys.   
The change in the trajectories for tiny changes in $V(0)$ observed is typical for chaotic motion
in general, and is a sign that a {\em hyperbolic invariant set} likely exists in the phase space
for the breakout motion of $P_3$. A hyperbolic invariant set in general is a Cantor set all of whose points are
have a saddle-like structure produced by a transverse homoclinic orbit, whose existence
is given by the Smale-Birkhoff theorem (Belbruno 2004). The use of the term {\em chaotic}
in a strict mathematical sense means the existence of a hyperbolic invariant set. 

It is
remarked that $Cl$ represents a {\em physical collision} with the surface of the Earth, where, at
periapsis below the Earth's surface, $r_2=.00003$. It turns out that actual {\em pure
collisions} where $r_2 \sim 0$ to high precision are readily found as well. For example, $V(0)=.011998$ 
leads to a pure collision at $t=360.1898$. 
In this type of collision, $P_3$ asymptotically approaches the collision manifold which is a set
of measure zero.

It is observed that since the motion of
$P_3$ repeatably passes near to the Earth in breakout motion, the Earth tends to 
readily pull $P_3$ toward pure and physical collisions. The set, or manifold,  of pure collision
trajectories are a subset of physical collision trajectories, and, in fact, are a 
set of measure zero in the four-dimensional phase space of position and velocity (Belbruno
2004). Since they are a set of measure zero, their near occurrence is reflective of the fact the
fly-bys of the Earth are close and that the Earth has a considerable gravitational focusing effect when the
trajectory is near parabolic.

\begin{figure}[ht!]
\begin{center}
\resizebox{70mm}{!}
  {\includegraphics{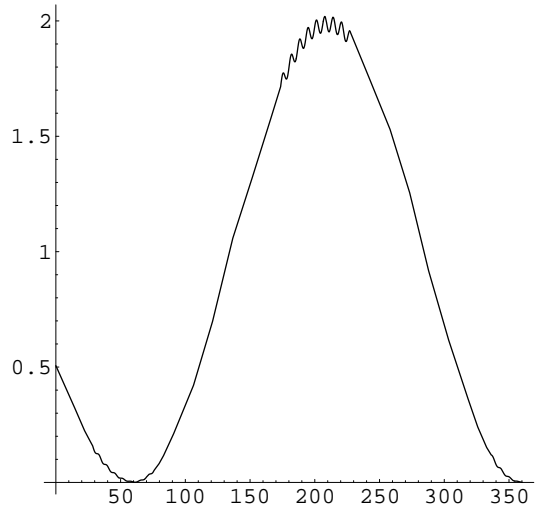}}
\end{center}
\caption{Kepler energy $E_2$ with respect to $P_2$ as a function of $t$
along entire
collision orbit. It is seen that $E_2 \rightarrow 0$.}
\label{Figure:ExtendedEnergy}
\end{figure}

\begin{figure}[ht!]
\begin{center}
\resizebox{90mm}{!}
  {\includegraphics{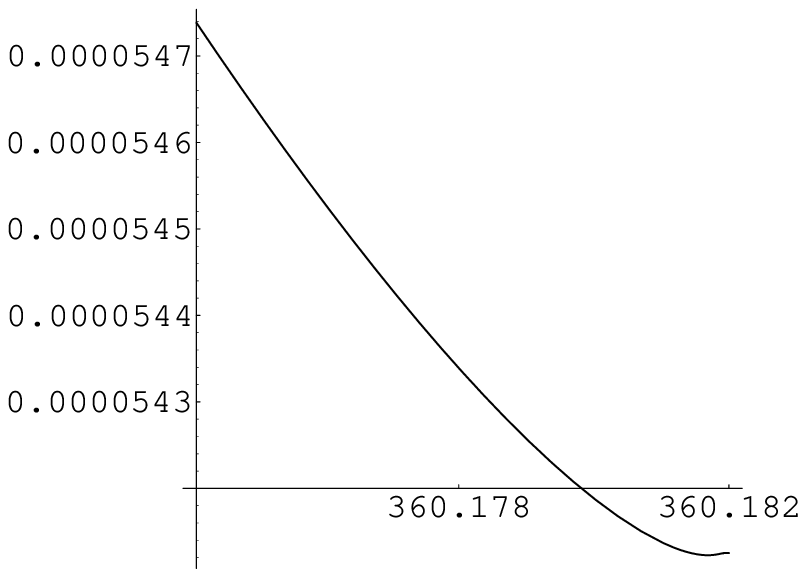}}
\end{center}
\caption{$E_2$ as a function of $t$ at the end of the collision orbit
approaching the value of .0000541 at periapsis below the earth's
surface, where
$r_2=.00003$ (i.e. 4488
km), for $t=.00014$ (i.e. 12 minutes) beyond earth collision, $t \in
[360.175, 360.1817]$. }
\label{Figure:EnergyAtZero}
\end{figure}

It turns out that $Cl$ is approximately parabolic at collision. This is seen by plotting the 
Kepler energy $E_2$ of $Cl$ with respect to the Earth. In inertial Earth centered coordinates, 
${\bold X} = (X_1,X_2)$, $ E_2 =  {1\over2}|{\bold{\dot{X}}}|^2 - 
{\mu\over{|{\bold X}|}}$. In
barycentric rotating coordinates ${\bf x} = (x_1,x_2) \equiv (x,y)$, 
$E_2$ is transformed into
\begin{equation}
 E_2 = {1\over2}|{\bold{\dot{x}}}|^2  - {\mu\over{r_2}} + 
       {1\over2}r_2^2 - L,
\label{eq:E2bary}
\end{equation}
where,  $ r_2 = \sqrt{(x_1+1-\mu)^2+x_2^2}, \ \ L = \dot{x_1}x_2-\dot{x_2}(x_1+1-\mu)$,  \ \ 
(Belbruno 2004, 2002). $L$ is the angular momentum of $P_3$.
$E_2$, (\ref{eq:E2bary}), is evaluated along 
$Cl$ and plotted in Figure \ref{Figure:ExtendedEnergy} as a function of $t \in
[0,360.181558]$. From Figure \ref{Figure:EnergyAtZero}, $E_2 = .000054$ at
collision, which is nearly parabolic. 


\noindent
(If the collision were purely parabolic then $E_2 = 0$.)
This yields a very slight hyperbolicity whose hyperbolic excess velocity with respect
to the Earth, $V_\infty = \sqrt{2E_2}$,
has the value of .0104. This is equal to $V^*(0)$ for $\alpha = \pi/2$ to within .0014. 
In scaled coordinates, $V_\infty = 310 m/s$ and $V^*(0) =  328 m/s$. $V_\infty$ is close
to $V^*(0)$ because at Earth periapsis $V^*(0)$ is like the velocity at infinity. This velocity is
approximately maintained along the orbit as it approaches collision. At
actual Earth fly-by at periapsis
the velocity with respect to the Earth increases due to the attraction
of the Earth, and for this collision orbit $V = .44$ at Earth periapsis.

In (Belbruno $\&$ Gott 2004) it is analytically shown that  
in the critical or near critical breakout motion, all close Earth fly-bys, including collision
trajectories, are approximately parabolic at periapsis. For critical breakout trajectories, which start at
$L_4$ at time $t=0$, $V(0) = V^*(0)$. For near critical breakout motion we assume
that $V(0) \stackrel{>}{\sim} V^*(0)$. Notationally, $V(0) \stackrel{\geq}{\sim}  V^*(0)$
includes both of these cases.

We define the terms, 'close Earth
fly-by', 'approximately parabolic'. Let $\gamma(t)$ be a trajectory 
which performs a fly-by of the Earth, with a periapsis distance
$r_2$ at some time $t$. We say that this is a {\em close Earth fly-by}
if $r_2 \leq 100,000$ km, or in dimensionless coordinates, $r_2 \leq .000668$. 
The figure of 100,000 km is arbitrarily chosen since for weakly hyperbolic fly-bys 
of the Earth beyond this distance, the effect of an Earth gravity assist is
negligible. Physical collisions are included as close
fly-bys.  
\medskip

We use $E_2$ to determine the type of collision, which is computed at Earth periapsis.
So, in the case of physical collision at the Earth's surface, we propagate the trajectory
to Earth periapsis within the Earth. This point occurs at a very short time after physical collision which for
the case of $Cl$ is only 12 minutes. At the periapsis point $E_2$ is evaluated 
at the trajectory state of position
and velocity. If
$|E_2| \stackrel{>}{\sim} 0$, the collision or collision trajectory is called
{\em approximately parabolic}. It could be slightly elliptic, slightly hyperbolic or
purely parabolic. It turns out, as we
will see, that for break out or near breakout motion, the fly-bys are all 
approximately parabolic. The following result is obtained (For details see Belbruno $\&$ Gott
2004):
\medskip

\noindent {\it For the set of critical or near critical breakout velocities at $L_4$,  
the value of $E_2$ at the close Earth fly-bys at periapsis
has the value
\begin{equation}
E_2 \approx {1\over2} V(0)^2  \stackrel{\geq}{\sim} {1\over2} V^*(0)^2
\stackrel{>}{\sim} 0 ; 
\label{eq:theorem2}
\end{equation}  
That is, the close Earth fly-bys are approximately parabolic. This is
true for all the values of $\alpha$ except those values in small
neighborhoods of  $5.102\pi/8, 13.5\pi/8$ (see comment below).}
\medskip

This implies that a trajectory $\gamma(t)$ starting near critical breakout velocity at
$L_4$ for $t=t_0$ will satisfy equation \ref{eq:theorem2} for 
any future time $t > t_0$
corresponding to any close Earth flyby at periapsis.

More precisely, as shown in Belbruno $\&$ Gott 2004, for a trajectory starting at $L_4$,
at Earth periapsis on a close Earth fly-by at a distance $r_2 = \delta_2$, 
\begin{equation}
E_2 = {1\over2}V^2(0) \ + \mathcal{O}_0(\mu) + \mathcal{O}_1(\delta_1) +
\mathcal{O}_2(\delta_2)
\label{eq:lemma3}
\end{equation}
where $ \mathcal{O}_0 = 3\mu, \ \mathcal{O}_1 = 2\delta_1, \ 
\mathcal{O}_2 = 2\delta_2 \cos \theta_2$,           
$r_2 = \delta_2 < .000668 \ll 1, \ r_1 = 1 + \delta_1, \
|\delta_1| < .000668, \ |\delta_1| \geq 0, \ \delta_2 \geq 0.$  This relation yields 
(\ref{eq:theorem2}) for $\mu, \delta_1, \delta_1$ small.

Let $P_3$ be at $L_4$(or $L_5$) at $t=0$, and let ${\bf v}(0)$ be the initial
velocity with magnitude $V(0)=|{\bf v}(0)|$. Then, the Jacobi integral
$J$ has the value $J = C_0 = 3 \ - V^2(0) \ -\mu(1-\mu)$.
This implies that for the set of critical breakout velocities $V^*(0)$ at $L_4$ for
$\alpha \in
[0,2\pi]$ (see Figure \ref{Figure:ScatteringDirections}), $C_0  \stackrel{<}{\sim} 3$.
\medskip

As mentioned earlier, there are two sharp spikes in the breakout velocities 
shown in Figure \ref{Figure:ScatteringDirections} of 
values .22, .25 which occur for $\alpha = 5.102\pi/8, 13.5
\pi/8$, respectively. However,  most values of $V^*(0)$ vary
between approximately $.006$ and $.05$ if two intervals in $\alpha$
of total width approximately $.157$ radians are deleted near where the spikes occur. 
This implies that for nearly all of the 
values of $\alpha$, equation (\ref{eq:theorem2}) implies that
approximately
\begin{equation}
E_2 \in [.000018, .0012]  .
\label{eq:E2para}
\end{equation}
Thus, $P_3$ is approximately parabolic at collision.
The range given by (\ref{eq:E2para}) is a crude estimate of $E_2$. The observed value for 
$Cl$ of $E_2 = .000054$ is contained within this interval.
A sharper estimate can be made using (\ref{eq:lemma3}). 
For $Cl$, $V(0)=.0119981$, and at fly-by periapsis, below the
Earth's surface, $r_2 = \delta_2 =.00003$. This occurs approximately
on the $x_1-$axis implying $r_1 = -\delta_1 = -.00003 , 
\ \theta_2 = 0$. 
Substitution of $V(0), r_1, r_2, \theta_2$ into (\ref{eq:lemma3})
yields the value, $E_2 \approx .00007$. Noting that the numerically observed value of $E_2$ 
at periapsis for $Cl$
$E_2=.000054$, the predicted value is in error by only .000016
demonstrating the accuracy of the predicted values.

\medskip

We remark that in all the cases numerically observed, $E_2$ was not
negative(i.e. the orbit was not elliptic) during the fly-by. Some such fly-bys are likely to 
exist due to the chaotic
nature of breakout motion, however, the probability of finding
trajectories with elliptic fly-by states is apparently small.
Their chaotic structure and low probability of occurrence is studied in 
Belbruno (2004).
When $E_2$ is near to zero this defines weak capture
studied in Belbruno (2004), where $P_3$ will, in general, move about
the Earth in a chaotic fashion generally leading to escape or collision. However, as remarked, the case of
interest here is when $E_2$ is very slightly hyperbolic.
\medskip

The retrograde collision trajectory $Cl$ emanating from $L_4$ 
is paired with another symmetric collision trajectory
$Cl^*$ emanating from $L_5$ which is 
symmetrical to $Cl$ and moves in a posigrade fashion about the Sun. This follows by 
the symmetry of solutions mentioned in Section \ref{Section:2}. It will collide with the
Earth in the 4th quadrant as shown in Figure \ref{Figure:Collision}. 

\medskip
\noindent
{\em Probability of Collision at Breakout for the Restricted Problem}
\medskip

A measure of the likelihood of finding collision trajectories is now described. 
This is done for the four basic initial velocity directions at $L_4$:
$\alpha = 0, \pi/2, \pi, 3\pi/2$. For an initial velocity of $P_3$
at $L_4$ for a given $\alpha$ we assume the corresponding 
breakout velocity $V^*(0)$ as shown in Figure \ref{Figure:ScatteringDirections}. The orbit of
$P_3$ is propagated from $L_4$ for $t \geq 0$ and since it is in
breakout motion we know that it will not be in horseshoe motion, but
will cycle about the Sun and repeatably fly past the Earth. We can
numerically demonstrate that collision with the Earth is likely.
This intuitively makes sense since the fly-bys will
be close and the the annular region supporting the breakout motion is
narrow. Now, for a given initial velocity at $L_4$ for $t=0$ we see from
Figure \ref{Figure:ScatteringDirections} that $V^*(0)$ is given up to three digits. For a given
value of $V^*(0)$, depending on $\alpha$, we propagate the trajectory
for up to $t=4000$, which corresponds to 637 years, and see if collision
has occurred. $t=4000$ is chosen arbitrarily, for convenience and is fairly small in
astronomical terms. If no collision occurred in that time, then we give
$V^*(0)$ a {\it random} perturbation by adding to it the random number
$.000mn$, where $m,n$ are positive random integers ranging from 0 to
9. For a choice of $m,n$ the trajectory is propagated again. If
collision does not occur, we repeat the process again for a different
choice of $m,n$, continuing trials until success is achieved. 

For $\alpha = 0, \pi/2$, we required two random trials
for success, where success means we achieve collision within 
$t=4000=637$years. For $\alpha =
\pi$, three random trials were required until we have success, and for
$\alpha = 3\pi/2$, six random trials were required for success.
Therefore we have achieved success in four random trials out of
thirteen. This gives our best estimate of the probability $\mathcal{P}$ 
of success for $0 \leq t \leq 4000$  as
\begin{equation}
\mathcal{P} \sim {4\over13}  .
\label{eq:prob1} 
\end{equation}
If we had not limited ourselves to $t=4000$ the probability would have been
larger. This probability is discussed in further detail in the Appendix.
We have run a sufficient number of trials to produce a rough order of magnitude 
estimate of this probability which is sufficient for our purposes, but 
a large number of additional trials could establish this number to higher
accuracy.

It is noted that the gravitational focusing on $P_3$ to cause a
collision is substantial. This is related to the fact the breakout
motion is occurring at a fixed energy for the planar restricted problem.
The fixed energy yields a three-dimensional energy surface obtained from
the Jacobi integral. As is proven in Belbruno (2004), the manifolds
leading to collision at $P_2$ are two-dimensional, and although they are
a set of measure zero, the particle $P_3$ is readily able to move
asymptotically close to these surfaces and to collision after the
gravitational focusing.  The collision manifolds on the Jacobi integral
surface separate the phase space, so it is fairly easy for $P_3$ to 
get near to the collision manifold. In higher dimensions this separation 
of the phase space on the Jacobi surface does not
occur, and the collision manifold is more elusive.

This concludes the demonstration of R2.
\medskip\medskip

It is interesting that these creeping chaotic orbits seem to lead
naturally to collision with Earth (as proposed by the giant impactor
theory) rather than to capture into a bound orbit (as in the sister
planet theory). If one wanted to have a sister planet theory, of course,
$L_4$ would be a promising place for the Moon to start out. So it is
significant that our chaotic creeping orbits (slightly hyperbolic-nearly
parabolic) lead naturally to collision rather than capture. This favors
the giant impactor theory. The sister planet theory would of course also
have a problem with the difference in iron between the Earth and the
Moon.

\medskip

\noindent
{\em Random Walk, $\Delta V$ Accumulation, and Relevant $V(0)$ Range} 
\medskip

In determining $V^*(0)$ at $L_4$ above, we kept $P_3$ fixed at $L_4$ and gradually
increased $V(0)$ for a given velocity direction. This yields a well defined 
set of $V^*(0)(\alpha)$ for $\alpha \in [0, 2\pi]$. 

We now consider a more realistic way that $P_3$ would increase its velocity in a gradual
fashion. The mechanism for this is to assume that $P_3$ is randomly being perturbed by encountering
other planetesimals(whether by gravitational encounter or direct collision) and in each 
encounter, it acquires an instantaneous kick $\Delta V$.
So, it is not kept fixed at $L_4$. To make this more realistic, we assume that the times
of encounters are random, within a large range, and the direction
$\alpha$ of the kicks are
random. The only thing we normalize is the magnitude of the $|\Delta V|$'s
which for convenience is held fixed. 

Thus, $P_3$ starts at $L_4$ with a zero velocity, and at time $t=t_1=0$ a velocity
$V(0) = \Delta V$, is applied in a random direction. This yields a vector 
${\bold v_1}$ with magnitude $\Delta V$. 
$P_3$ moves on a trajectory $\gamma(t)$ in a 
neighborhood of $L_4$, assuming that the value of $\Delta V$ is small. At a random time
$t_2 > 0$ another velocity
vector ${\bold v}_2$ of random direction and magnitude $\Delta V$  is vectorially added to $P_3$'s
velocity  at $t = t_2$. Then the trajectory is propagated for $t > t_2$ until at
another random time $t_3 > t_2$ a random vector ${\bold v_2}$ of magnitude
is vectorially
added to  $P_3$'s velocity vector at $t=t_3$, and this process continues creating a sequence
$t_k$ of times, $t_{k+1} > t_k$, and velocities ${\bold v}_k, k=1,2,3, \ldots.$

While the $\Delta V$'s are being applied, the trajectory $\gamma(t)$ is gradually moving further from
$L_4$, but since the velocity directions ${\bold v_k}$ are applied randomly, the 
path of the trajectory $\gamma(t)$ will move further away from $L_4$ for some time spans, and then
move toward $L_4$ for others. However, as $k$ increases, one would
expect, by the 
principle of random walk, for $P_3$ to eventually escape $L_4$ and creep toward the
Earth for $k$ sufficiently large when the  velocities ${\bold v}_k, k=1,2,3, \ldots$ 
applied on the trajectory $\gamma(t_k)$, gradually accumulate to a sufficiently large magnitude for breakout
to occur.  
If the ${\bold v}_k$ were all applied in the direction of motion of $P_3$ at $t_k$, then
the magnitudes $\Delta V$ would add producing an cumulative velocity addition of
$k\Delta V$ at the $k$th step. However, the directions of ${\bold v}_k$ are random, and 
by the principle of a random walk, the number of encounters $k$ before ejection occurs 
should be expected to instead satisfy 
\begin{equation}
\sqrt{k}\Delta V \approx .006 
\label{eq:randomwalk}
\end{equation}
for $k$ sufficiently large(and $\Delta V$ sufficiently small) where 
$.006$ is approximately the minimum value .0057 of $\{ V^*(0) \}$.
This makes dynamical sense, since
as the $\Delta V$'s are applied, the trajectory $\gamma(t)$ would seek to minimize the Jacobi energy, and hence
the velocity,
along its path. We found in all our numerical simulations the following result:
\medskip 

\noindent {\it For a given value of $\Delta V$, the number $k$ of random ${\bf v}_k$
applications required for breakout to occur approximately satisfies (\ref{eq:randomwalk})}.

\medskip

We describe this process of {\it random walk $\Delta V$ accumulation},
and its verification.

\medskip

For convenience we choose $\Delta V = .001$ and start at $L_4$ with zero initial velocity.
(\ref{eq:randomwalk}) implies that $k$ should satisfy $\sqrt{k}\Delta V \sim .006$,
which yields $k \cong 36$.  It was found that breakout occurred when $k=36$ as predicted.
Random kicks of velocity $\Delta V$ were applied in random directions $\alpha_k$ and 
after random time intervals $t_k$. 
(As in all future runs the times between kicks are just chosen to be large enough to randomize the
position. The real time for random walkout is expected to be much longer in years - perhaps
30 million years as considered in the Appendix.)  

From the above, we have the following result,
\medskip

\noindent
(Random Walk $\Delta V$ Accumulation) \ \ Under a realistic assumption of random walk, the 
peculiar velocity for $P_3$ accumulates proportional to the square root of the number of encounters
until it reaches a breakout state. Since a random walk
is isotropic the peculiar velocity is likely to encounter the breakout state first at a point
near the minimum value of .006 of the set $\{V^*(0)\}$ thus giving (\ref{eq:randomwalk}).  
\medskip

Therefore, substituting $V^*(0) = .006$ into (\ref{eq:theorem2}) 
implies that for close Earth fly-bys resulting from the random walk process at or near breakout,
$E_2 \approx .000018$. This implies that at close Earth fly-by resulting from the random walk process, a nominal value 
of $V_\infty = .006$ which is 179 m/s. 
When $P_3$ does a close fly-by of the Earth, after passage through periapsis it will
receive a gravity assist and increase, or decrease, its velocity with respect to the Sun.
A measure of this velocity change is observed due to the bending of the trajectory
of $P_3$ as it passes through periapsis. For example, this bending is clearly seen in Figure
\ref{Figure:Pass2}. The more distant the fly-by, then in general the less the bending.
The maximum bending is obtained from pure collision trajectories, where the bending angle is
$\pm\pi$. It is determined in Broucke (1994) that the resulting change in magnitude of 
the velocity, $\delta v$ with
respect to the Sun due to gravity assist is maximally, $2V_\infty$. Thus, {\it for each close
Earth fly-by, the expected maximum gain in velocity magnitude is approximately 358 m/s}. In general,
they will be less.  

The maximal velocity of 358 m/s is a relatively small number and will have little effect on a 
breakout trajectory when it has a close Earth fly-by. This velocity 
is less than .012 of the orbital velocity, inducing eccentricities into the trajectory
of $P_3$ after fly-bys of at most this order.
It is found in general that 
within time spans on the order of 2000 time units, there
are generally only one or two close Earth fly-bys. This implies that {\it $P_3$ will remain in 
breakout motion about the Sun in a relatively thin annular region for very long periods of
time, generally tens of thousands of time units, and repeatably pass by the Earth 
without being ejected}.

\noindent
{\em Collisions in Three Dimensions and the Mars Sized Impactor}
\medskip

Thus far we have constrained $P_3$ to lie in the plane of motion of
the Earth and Sun in the planar restricted three-body problem. 
It that situation it was seen that collision trajectories 
are readily found. When the motion of $P_3$ has an out of plane component, $z$, added to
it, it is more complicated. In this case we have the three-dimensional circular restricted
three-body problem, defined exactly the same way as for the planar problem, except that $P_3$ can move
in the $z$ direction, see (Szebehely 69; Belbruno 2004). By continuity with respect to initial
conditions, it must be the case that collisions in the planar case will persist in the three-dimensional  
case if $|\dot{z}|$ is sufficiently small.  
A way to approximately get a measure of the maximal allowed $z$ motion is
to consider a collision trajectory from $L_4$ to the Earth generated at
or near critical breakout, and then see how much $\dot{z}$ can be added
at $L_4$ and still maintain collision with the Earth which is now a
three-dimensional sphere. This is a relatively
straight forward calculation. 

In the Appendix we show that Earth collisions should persist providing
$|\dot{z}| \stackrel{<}{\sim} .0034$. This implies a thin disk of planetesimals.
In the Appendix we show that a disk of thickness $28 r_E$ easily satisfies this 
requirement. This has an angular width of 4.1 minutes of arc as seen from the Sun.
Although this seems thin, it turns out that inner B, A rings of Saturn,
extending from 92,000 km to 140,210 km, have a thickness of .18 - 1.7 
seconds of arc as seen from the center of Saturn, which is even thinner than required in our situation.

The modeling of most interest in this paper is for the general three-dimensional three-body problem defined by
equation \ref{eq:5}, where $m_3 = .1 m_2 \neq 0$, so a Mars-sized Earth
impactor is modeled. This is used by Canup $\&$ Asphaug (2001).
Although the motion of the Earth,
$P_2$, is given initial conditions for uniform circular motion, about the Sun, $P_1$,
it need not remain circular as time progresses due to the gravitational perturbations of $P_3$.   
This property makes the problem more interesting. To better understand this and to see its effect on
collision trajectories, the general planar three-body problem is first considered. It is defined from
(\ref{eq:5}) by setting $X_{k3}=0, k=1,2,3$. 

As with the restricted problem, a rotating coordinate system $(x,y)$ is chosen, this time, rotating with the
mean motion of the Earth about the Sun. In this system the Earth is not fixed on the -$x$ axis as in
the restricted problem since it is perturbed by $P_3$ especially during fly-bys.

Just as in the restricted three-body problem, breakout from $L_4$ can be
defined for the general planar three-body problem with exactly the same
methodology as for the restricted problem, by gradually increasing the velocity of $P_3$ at
$L_4$ until breakout is achieved, for any given velocity direction. 
Nearly identical results are obtained as in R1(Figure
\ref{Figure:ScatteringDirections})
and this analysis is not duplicated for this situation. 
The desired random walk $\Delta V$ accumulation process is defined exactly as
before for this current problem, and we have verified that the same results are
obtained.

We determine the
breakout of $P_3$ to Earth collision as we did in the restricted problem
by gradually increasing the velocity at $L_4$ until a critical velocity
is reached. That is, we are not modeling the random walk process, and are copying
the procedure we performed for the restricted problem. This is done to keep the modeling as close as
possible to the restricted problem in order to better understand how $P_3$ perturbs $P_2$ and the
effect this has on obtaining collision trajectories. 

\begin{figure}[ht!]
\begin{center}
\resizebox{70mm}{!}
  {\includegraphics{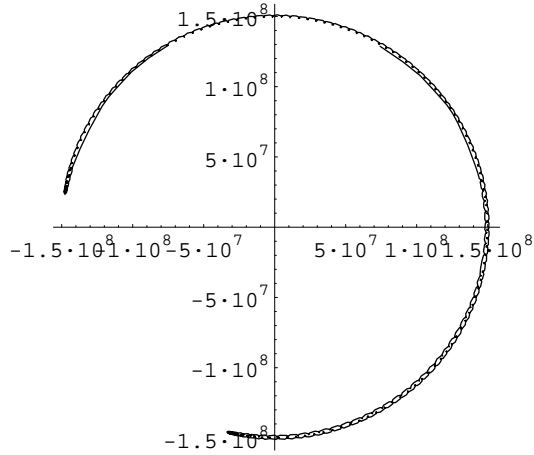}}
\end{center}
\caption{$\gamma(t), V(0) = .160$ km/s, $t\in [0,200] years$, x vs y,
Sun centered.}
\label{Figure:horse3}
\end{figure}

An important difference in using the planar three-body
problem instead of the restricted problem is observed when the breakout velocity
is determined for $P_3$ at $L_4$. As the velocity
magnitude $V(0)$ for $t=0$ is gradually increased at $L_4$, and 
$P_3$ moves in the horseshoe regions about the Sun, the location of the 
Earth moves in its orbit, approximately maintaining its 1 A.U. distance
from the Sun, but shifting its angular position with respect to the Sun.
This is because as $P_3$ creeps further and further away from $L_4$
approximately on Earth's orbit, it can gravitationally perturb the Earth
when it moves relatively near to the Earth since its mass is now one tenth
that of Earth. Analogous to the restricted problem, when $V(0)$ gets close to 
the breakout value, the horseshoe region begins to close on itself in a
symmetrical way with respect to the Earth. However, since the Earth has
shifted its location, the symmetrical closing point is not on the $x$-axis
as in Figure \ref{Figure:V009} for the restricted problem, but at another
location, the Earth's location, approximately 1AU from the Sun.  
This is illustrated in Figure \ref{Figure:horse3} where  
$V(0) \equiv | (\dot{x}, \dot{y}) | = .160$ km/s. $\alpha = 0$ is assumed. 
This is fairly close to
breakout, for the given velocity direction,  which occurs for $V(0) = .205$ km/s. The final location of the
Earth when breakout is achieved is in the third quadrant about 20 degrees
from the $-y$-axis which means that the $x$-axis is not fixed to the
Earth but to the mean motion.

The cases we have examined indicate that the likelihood for collision
to occur in this problem is similar to that of the restricted problem.
However, the dynamics of collision is more complicated. 

We now examine a collision trajectory, $Cl2$ and the associated dynamics.
$Cl2$ occurs in the breakout state for $V(0)=205$ km/s. 
This is only discussed briefly here as the details can be found in Belbruno $\&$ Gott (2004).
It starts at $L_4$ for $t=0$ with a velocity of .205 km/s in the positive $x$-direction. 
The Earth is located initially on the negative $x$-axis at 1AU distance from the Sun, at the origin.

\begin{figure}[ht!]
\begin{center}
\resizebox{70mm}{!}
  {\includegraphics{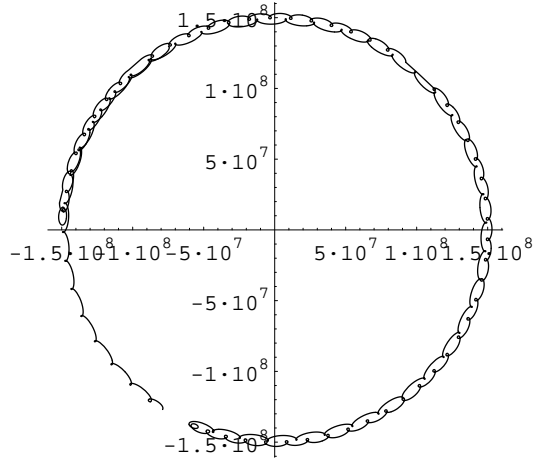}}
\end{center}
\caption{$CL2$ from $L_4$ at $t=0$ to earth collision at t = 108.628235
years in the
third quadrant. Sun centered, x vs y.}
\label{Figure:fullcollision3}
\end{figure}

$Cl2$  is plotted in Figure \ref{Figure:fullcollision3}

In Figure \ref{Figure:fullcollision3} in the rotating 
coordinate system $P3$ escapes $L_4$ moves down toward the Earth, then turns around and moves  
in a retrograde motion (clockwise)
about the Sun, continuing to the third quadrant, and then in a {\em
posigrade} motion (counterclockwise) with respect to the Sun, it circles the Sun, moves past the negative 
$x$-axis, to Earth collision. 
Unlike $Cl \equiv Cl1$, $Cl2$ is a posigrade collision orbit. 

\begin{figure}[ht!]
\begin{center}
\resizebox{70mm}{!}
  {\includegraphics{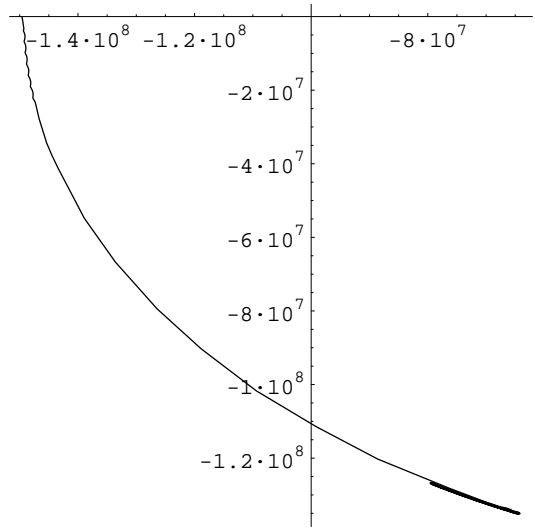}}
\end{center}
\caption{The motion of the earth from $t=0$ to collision with $P_3$ on
$Cl2$ at t = 108.628235
years,  x vs y.}
\label{Figure:shiftedearth}
\end{figure}

Now, as $P_3$ has moved in this collision orbit, the Earth has moved also. It has moved
in the posigrade direction and then, toward the end in a complicated
motion in the retrograde direction. This is seen in Figure \ref{Figure:shiftedearth}

\begin{figure}[ht!]
\begin{center}
\resizebox{70mm}{!}
  {\includegraphics{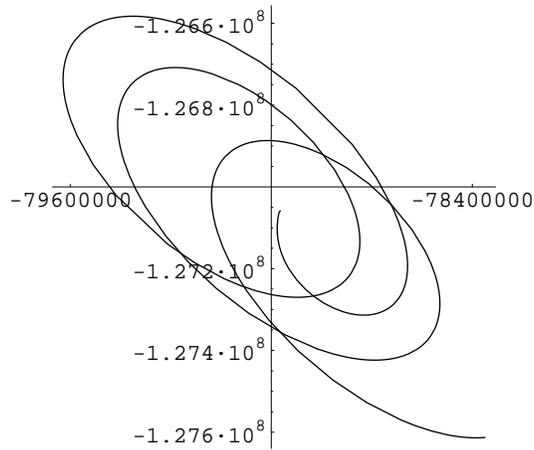}}
\end{center}
\caption{The motion of the earth(note enlarged scale) about 4 years
prior collision
with $P_3$ on $Cl2$ at t =
108.628235years. x vs y.}
\label{Figure:complicatedearth}
\end{figure}

The motion of the Earth about 30 years prior to collision with $P_3$ is complicated.
The final phase of the Earth's motion much enlarged is shown in
Figure \ref{Figure:complicatedearth}. It consists of many small
loops, one for each year, caused by the perturbation of $P_3$ as it gets near to
collision. In Figure \ref{Figure:complicatedearth}, this looping motion is shown
in large scale about four years prior to collision with $P_3$. 
The Earth is moving in a retrograde fashion in this figure,
starting in the lower right and ending in collision near the center of
the coordinate system. The very end of the Earth's trajectory is
actually parabolic in appearance, as seen in Figure
\ref{Figure:actualcollision3big}, which is too small to be seen in
Figure \ref{Figure:complicatedearth}.

\begin{figure}[ht!]
\begin{center}
\resizebox{70mm}{!}
  {\includegraphics{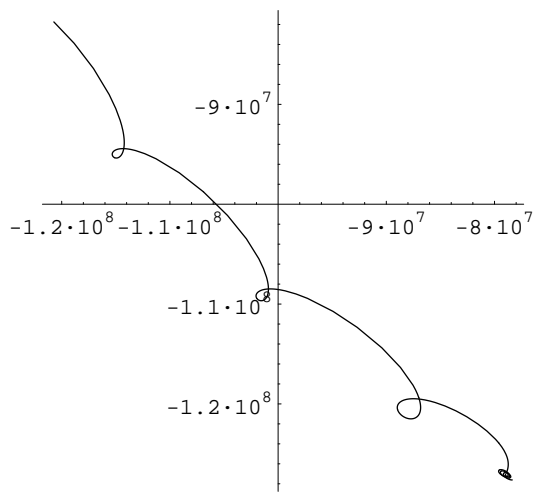}}
\end{center}
\caption{$Cl2$ 4 years prior collision with the earth at
t = 108.628235years.  x vs y.}
\label{Figure:complicatedorbit}
\end{figure}

\begin{figure}[ht!]
\begin{center}
\resizebox{90mm}{!}
  {\includegraphics{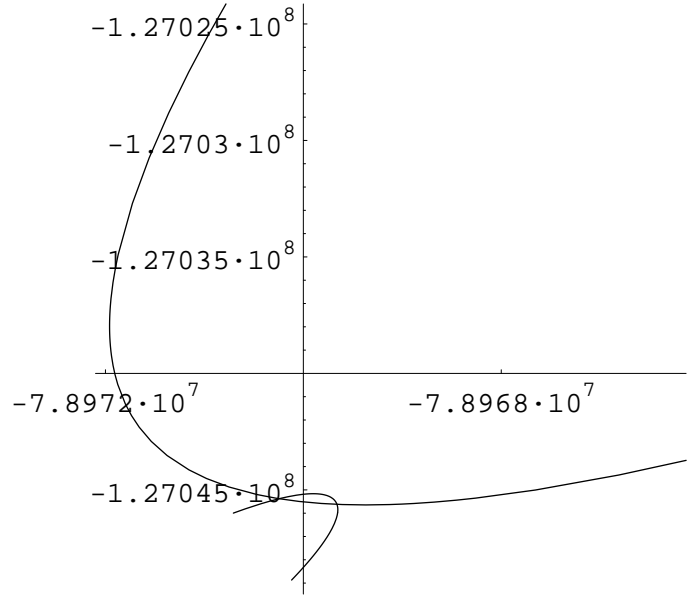}}
\end{center}
\caption{Relative motions of the earth and $P_3$ near collision.
$CL2$ moves on the larger parabolic curve clockwise from right to left,
and the earth moves on the smaller parabolic curve clockwise from left
to right. Time Duration: 1.9 hours. x vs y.}
\label{Figure:actualcollision3big}
\end{figure}

In the same time frame as in Figure \ref{Figure:complicatedearth}, we show
$P_3$ on $Cl2$ moving to collision with the Earth in
Figure \ref{Figure:complicatedorbit}. The small black smudge
on the lower left is the complicated motion of $P_2$(shown enlarged in Figure
\ref{Figure:complicatedearth}) prior to collision
which relative to the scale of $Cl2$ is too small to be clearly seen.

The actual Earth collision with $P_3$ is shown in Figure \ref{Figure:actualcollision3big}
which shows the relative motions of the Earth and $P_3$. We have integrated 
the motion of $P_3$ through collision to better show the relative motions.
In this figure, $Cl2$ is the larger parabolic type curve, and the Earth moves in the
smaller curve. The Earth moves clockwise from the left to the right, and
$P_3$ moves clockwise from the right to the left. The span of the
vertical axis is approximately 20000 km, so that one half of this
distance represents the radii of the Earth and $P_3$ added together.
This implies that actual physical collision between the Earth and $P_3$
occurs in this figure when $P_3$ is near the start of its motion on the
bottom right. 
We have verified that this is a near parabolic
collision as in $Cl1$. 

\medskip
\noindent
{\em Three-Dimensional Simulation of Collision in an Anisotropic
Thin Planetesimal Disk via Random Walk Encounter Dynamics}
\medskip

We now consider the main model of this paper for the numerical simulation of a collision
trajectory by a Mars-sized impactor. So, the planar three-body problem just considered 
is now generalized to three-dimensions given by (\ref{eq:5}). At $t=0$ the Earth has the same initial position 
position as in the planar case, and $P_3$ starts at $L_4$. We now more realistically model the 
breakout of $P_3$ from $L_4$ by the random walk $\Delta V$ accumulation
process.

We assume that $\Delta V$'s are imparted to $P_3$ at random
times. We will also assume that at these random times separate
independent $\Delta V$'s are
imparted to the Earth. As is described in the Appendix, we are assuming
a thin planetesimal disk, and at the random times, two types
of velocity kicks are applied to both the Earth and $P_3$. 
One velocity kick is assumed to be in a random direction in the plane, and labeled $\delta V_{\|}$,
and the other is perpendicular to the plane randomly either up or down and labeled $\delta V_{\perp}$.
The velocity kicks applied to the Earth have a subscript of E, and those
relative to $P_3$ have no subscript. In the 
Appendix we estimate the magnitudes of these velocities. The
magnitudes for the velocity kicks for $P_3$ are given by (\ref{eq:A7}) in the Appendix, and the magnitudes for
the Earth are given by (\ref{eq:A6}) in the Appendix. 
In this case breakout occurs when $k=14$.  
After these directions are input, the trajectory is propagated for a random
time $t$ which varies between 0 and 100 years. The process is then repeated.

Once the breakout state was achieved in the 14th step, and the propagation terminated
after $t_{14}=93.1$ years, it was found that by extending it another 6.9 years to
100 years, no collision occurred. We then went back to the 
beginning of the 14th step. A different set of random
values were given to the velocity kick directions for both the Earth and $P_3$,
keeping the time of integration to be from 0 to 100 years.  Collision again did not occur. 
We then again went back to the beginning of the 14th step, and again picked random values, and  
collision also did not occur. We then made a third additional random trial for the velocity kick directions
at the beginning of the 14th step and found collision did occur
when $t_{14} \sim 4.56$ years.

This yields a probability of collision $\mathcal{P}=
1/4$(because in one of four random trials we succeeded) we
discuss this further in the Appendix. (As in the discussion following equation \ref{eq:prob1},
our number of trials is sufficient to give a rough estimate of this probability
which is all we require, and additional trials could establish this number to higher
accuracy.)  

\begin{figure}[ht!]
\begin{center}
\resizebox{80mm}{!}
  {\includegraphics{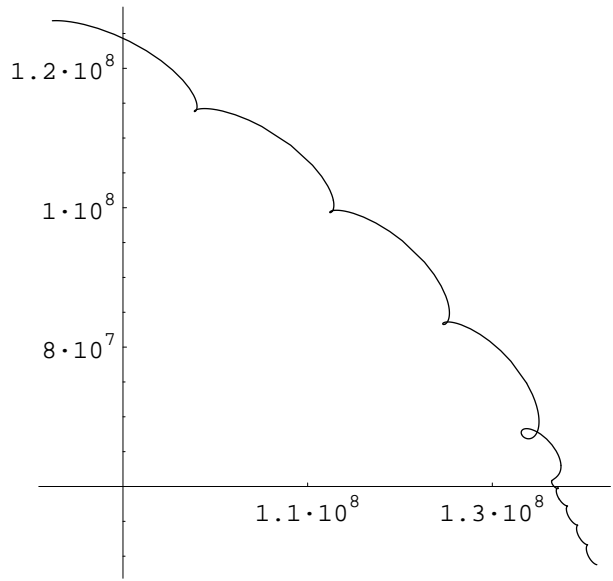}}
\end{center}
\caption{Collision between Earth(Lower curve) and Impactor(upper curve)
in
the first quadrant.
Earth moves in upward (posigrade) direction, Impactor moves in downward
(retrograde) direction.
Collision occurs slightly above the x-axis.
Time Duration: 4.5615948 years. Planar Projection, x vs y.}
\label{Figure:AnisotropicCollision}
\end{figure}

The collision trajectory $\gamma(t)$ is plotted in Figure
\ref{Figure:AnisotropicCollision} from the initial value in the
breakout state at the 14th step using the randomly chosen velocity
kick directions in the final attempt.
The time from its initial condition is 4.5615948 years, and
this final portion of the trajectory $\gamma(t)$ is shown. In Figure
\ref{Figure:AnisotropicCollision} it is the upper curve, and the motion
is in the downward direction. The smaller lower curve shows the motion
of the Earth which moves in the upward direction. The collision is seen
to take place near the $x-$axis(when the center of $P_3$ hits the Earth's surface).
Actual physical collision occurs slightly earlier when the surface
of the impactor hit the surface of the Earth. If we continue the trajectory $\gamma(t)$
through the Earth's surface to Earth periapsis, the periapsis distance 
is only approximately 200 km. The time of periapsis is 4.5616056 years.
See Belbruno $\&$ Gott(2004) for more details.

The initial conditions for the Earth and $P_3$ at the 
beginning of breakout, 4.5615948 years prior to Earth collision are explicitly
given in \cite{BelbrunoGott}. This breakout state results from the random walk process previously described after 13 velocity kicks at
times $t_i, i=1,2,..., 13$, where $t_{14}=4.5615948$ years. The total time
$T$ for the motion of $P_3$ to reach this state is
$T = \Sigma^{14}_{i=1}t_i = 728.2$ years.
The position of the Earth at breakout shows that the Earth has migrated in its
orbit a considerable distance in a posigrade fashion from its initial position on the negative
x-axis to the first (upper right) quadrant approximately along its orbit.
The velocity of the Earth has slightly changed to 29.773 km/s from 29.78 km/s due
to perturbations of $P_3$. Even though collision with the Earth is 4.56 years away,
the velocity of $P_3$ is 29.698 km/s which differs from that of the Earth by only
75 m/s.

We describe the collision trajectory of $P_3$ with the Earth. The coordinate system is the same that
we used in describing the motion of $Cl2$; that is, a rotating
coordinate system rotating with the mean motion of the Earth about the 
Sun. It is convenient use Jacobi
coordinates  ${\bf q} = (q_x, q_y, q_z) \ \ {\bf Q} = (Q_x, Q_y, Q_z)$, where
${\bf q}$ is the relative
vector of the Earth with respect to the Sun, and ${\bf Q}$ is the vector
from the center of mass of the binary pair $P_1,P_2$ to $P_3$ (Belbruno 2004).
As with the planar three-body problem we considered, we use a rotating
coordinate system which initially rotates in the plane of the Earth about 
the Sun, and with the mean motion of the Earth about the
Sun.  

\begin{figure}[ht!]
\begin{center}
\resizebox{70mm}{!}
  {\includegraphics{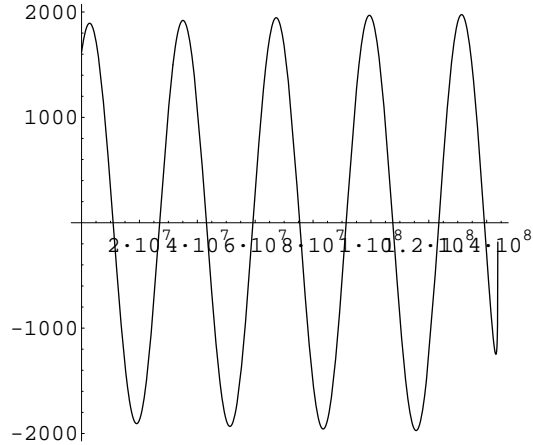}}
\end{center}
\caption{Variation of $q_z-Q_z$.
Time Duration: 4.5615948 years.}
\label{Figure:zoscillation}
\end{figure}

Along the collision trajectory $\gamma(t)$ of $P_3$ the $z$-variation between 
the Earth and $P_3$, given by
$q_z-Q_z$, oscillates between approximately $\pm 2000$ km. This is shown
in Figure \ref{Figure:zoscillation}. 

\begin{figure}[ht!]
\begin{center}
\resizebox{70mm}{!}
  {\includegraphics{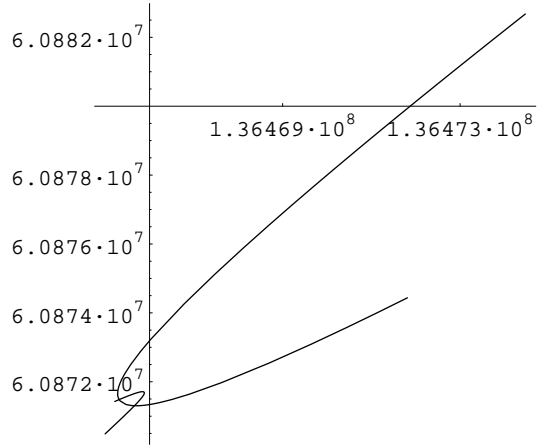}}
\end{center}
\caption{Relative motions of the Earth(smaller parabolic curve) and
Impactor(larger parabolic curve) near collision. Earth moves
counterclockwise from left to right, Impactor moves counterclockwise
from right to left. Time Duration: .53
hours. Planar Projection, x vs y.}
\label{Figure:AnisotropicCollision2}
\end{figure}

We show the relative motion of the Earth and $P_3$ near collision in
Figure \ref{Figure:AnisotropicCollision2}. In this figure the time
duration is only .53 hours. The orbits of the Earth and $P_3$ have been
continued beyond collision to get a better understanding of the
dynamics. The vertical axis spans approximately 10000 km,
so that actual collision of the surface of the impactor with the 
surface of the Earth would occur near the beginning of the trajectory $\gamma(t)$ in the 
upper right quadrant. 
This dynamics is analogous to that of $Cl2$ near collision
shown in Figure \ref{Figure:actualcollision3big}. 

As a final comment, it has been verified that including full solar
system modeling, as described at the end of Section \ref{Section:2},
the process of obtaining the collision trajectory $\gamma(t)$ is perturbed by
a negligible amount, and a nearby collision trajectory can be constructed.
This is due to the fact that at breakout, the motion of $P_3$ stays 
close to 1AU radial distance from the Sun. It is also noted that 
in our analysis of the motion of $P_3$ after breakout, it performs
several close flybys of the Earth for several hundred years. During this
time collision is fairly likely to occur. Our analysis has shown that it is
most likely to occur soon after breakout, which for the trajectory $\gamma(t)$ was
only 4.6 years. It has been found that after breakout has
occurred $P_3$ generally makes very close flybys of the Earth
for up to approximately 500 years, and then the flyby distance becomes
steadily larger. The repetitive flybys increase the
semi-major axis and eccentricity of the trajectory making collision 
less likely as $P_3$ is no longer restricted to staying as close to 1AU
radial distance from the Sun as it did immediately after breakout. Also,
$P_3$ will acquire larger deviations in the out of plane directions. Soon
after breakout, these effects are significantly less pronounced. In this
sense, soon after breakout $P_3$ has a good chance to creep into an
Earth collision; however, if it does not have one
relatively soon then the flybys themselves will eventually make
collision less likely. Overall, as we have shown, the probability of
collision with Earth rather promptly after breakout is appreciable (of
order 1/4). As discussed previously, running additional cases could establish 
this number to greater precision. 
\medskip

\noindent
\section{Discussion}
\label{ref:Conclusions}
\medskip
\medskip

We have shown that the giant impactor could have formed at $L_4$($L_5$)
and then escaped on a creeping chaotic trajectory to impact the Earth,
with a near parabolic encounter in agreement with simulations.

We note that there are difficulties if the giant impactor 
came from a location other than $L_4$(or $L_5$). To illustrate this 
assume that it came from elsewhere.

Since the Earth's orbit and Venus' are nearly circular and co-planar
even at the current epoch, after 4.5 billion years of perturbations,
this suggests that the early disk of planetesimals in the neighborhood
of the Earth was quite thin and that the planetesimals in the disk were
in orbits that had low eccentricity (e) and inclination (i), $e \sim i
\ll 1$. The critical impact parameter for collision with the Earth for
a small planetesimal is 
$$b_m = r_E(1+[V_{es}^2/V_{pec}^2])^{1/2},$$
where $V_{pec}$ is the peculiar velocity of the 
planetesimal, i.e., $V_{pec}\sim e \times V_{orb} \sim e \times 30$ km s$^{-1}$,
and $V_{es}$ is the escape velocity from the surface of the Earth (see Appendix).
If  $V_{pec}/V_{orb} < 0.004$, then $b_m \sim r_E(V_{es}/V_{pec})
> e \times 1AU \sim i \times 1AU \sim (V_{pec}/V_{orb})1AU$, and
the planetesimals whose semi-major axes are within a distance of 
$b_m$ of the Earth's distance from the Sun of 1AU will likely
suffer collision with the Earth within a short number of years since we
expect the orbits to be chaotic, and the impact parameter with the Earth
is less than the critical impact parameter $b_m$
for collision with the Earth. This will clear out a region of $\pm
b_m$ around 1AU {\it Except for planetesimals in stable orbits around
$L_4$(or$L_5$)}. Planetesimals at nearly 1AU from the Sun--and not
at $L_4$(or$L_5$)--will quickly be accreted by the Earth, before having
the chance to grow large by accretion themselves. T.R. Cowley (lecture at Univ. of Michigan 2002)
has noted this problem, saying that "Advocates of the Big Whack hypothesis usually say that
the impactor must have been formed near the Earth. This is neither
probable nor impossible. It is not probable because the Earth could have
readily swept up materials that would have formed the other body. It
is not impossible because we do not know the precise conditions of the
accumulation of the Earth, and cannot say how improbable assembly of the
putative impactor near one astronomical unit really was.")  The stable
location at $L_4$(or $L_5$) answers the probability question, offering a
reasonably likely scenario for forming the giant impactor near 1AU without
the material first being swept up by the Earth.  Once this cleared out
region of $\pm b_m$ has been established, there will be no further quick
accretion onto the Earth, because the planetesimal's orbits will not take
them to within an impact distance $b_m$ from the Earth. Then they will
have to diffuse in by two-body relaxation-from perturbations by other
planetesimals and planets. This two-body relaxation process will slowly
put planetesimals into the gap region again and there will be be quick
accretion from the gap.  The giant impactor is expected to be one of the
later impactors to hit the Earth because the successful simulations of the
formation of the Moon start with the Earth already at nearly its current mass,
showing that its subsequent accretion (after the giant impactor hit)
is assumed to be small (Canup $\&$ Asphaug 2001). The giant impactor should also be expected to
be one of the latter impacts because planetesimals, including the Earth,
grow by accretion with time and that would have also allowed more time for
the giant impactor to have grown by accretion itself. 

If the giant impactor is one of the {\it latter} impactors as argued by 
Canup $\&$ Asphaug (2001) (after most
accretion for the Earth has been completed) then if it is not from
$L_4$(or $L_5$) it must originally come from {\it either} significantly outside
1AU or significantly inside 1AU. But then it would violate one of
the key advantages of the great impactor theory: namely, point (3)
in the Introduction, which explains why the Earth and Moon have the
same oxygen isotope abundance- namely that the Earth and the giant impactor
came from the same radius in the solar nebula. Meteorites from
different neighborhoods in the solar nebula(those associated with parent bodies of Mars and Vesta for 
example) have different oxygen
isotope abundances. The impactor theory is able to explain the otherwise
paradoxical similarity between the oxygen isotope abundance in the Earth and the
Moon combined with the difference in iron.

The Earth has oxygen isotope abundances that are an average over all
the planetesimals it has accreted--some initially from inside 1AU and
some from outside. A giant impactor forming outside 1AU and drawn in by
two-body interactions would have oxygen isotope abundances intermediate
between Earth and Mars and therefore not identical with the Earth.
Standard giant impact theory has the Moon formed primarily out of 
mantle material from the giant impactor material (see (Canup 2004)).
The rest of the material in
the giant impactor is absorbed by the Earth and the iron core of the
giant impactor eventually finds its way into the Earth's core, leaving
the Moon iron depleted relative to the Earth. The Moon has been found
to have a small core and this is assumed to be from giant impactor
material. (See (W\"anke 1999) for a discussion of how the giant impactor 
theory can accommodate this.) 
If the Moon derives from giant impactor 
material then it would have, the theory proposes, isotopic abundances
identical with the Earth if it was formed near 1AU and this is observed to be the
case (Clayton $\&$ Mayeda 1996; Wiechert et al 2001; W\"anke 1999;
Lodders $\&$ Fegley 1997 . But if the giant
impactor came from significantly outside outside 1AU its isotopic
abundances would be significantly different from that of proto-Earth.
Furthermore, since $M_I$ is only  
10$\%$ the mass of the Earth, this would pollute the proto-Earth's isotopic abundances
with only a 10$\%$ contribution from the giant impactor.  This would give the
Earth and the Moon different isotopic abundances, if the giant impactor came
from significantly outside 1AU. A similar trouble occurs if the giant
impactor originated significantly inside 1AU, if the oxygen isotope abundances inside 1AU are heterogeneous
as well. (At present we have no meteorites in our possession whose parent bodies are thought to be
Mercury or Venus. So we currently have no data for oxygen abundances inside 1AU.)

On the other hand, consider what happens if the giant impactor originated
at $L_4$(or $L_5$). It is in a stable orbit, so it is not immediately
accreted onto the Earth, and can grow large and hit the Earth later,
alleviating the problem mentioned by Cowley.  It sits nicely at 1AU
and accretes exactly the same type of material the Earth does, some
diffusing from outside 1AU, and some from inside. The integral of the
oxygen isotope abundances of the accretion should be identical with that
of the Earth. Eventually, perturbations kick the giant impactor out
of its stable orbit and it collides quickly with the Earth. When the giant impactor hits the Earth and kicks out the
Moon, since the Earth and giant impactor have identical isotope ratios,
the Earth and Moon should have identical isotope abundances even though
the Earth and Moon are polluted to different extents by giant impactor
material. This is an advantage to the giant impactor model, producing
automatic agreement with proposition (3) of the giant
impactor model. Since this is one
of the latter accretion events for the Earth in terms of the accumulation
of its mass, the oxygen isotope abundances for the Earth and Moon will
not be further significantly changed by post-giant impactor accretion.

Thus we propose the following scenario.
\medskip

Debris remains at $L_4$ (as the Trojan asteroids prove). From this 
debris a giant impactor starts to grow like the Earth through accretion 
as described above. As the forming giant impactor reaches a sufficient 
mass ($\sim .1 m_{Earth}$), it 
gradually moves away from $L_4$ through gravitational encounters with other 
remaining planetesimals
and it randomly walks in peculiar velocity. It gradually
moves farther and farther from $L_4$ approximately on the Earth's orbit
in a horseshoe orbit at 1AU, until it acquires a peculiar velocity 
of approximately
180 m/s. The giant impactor then performs breakout motion where it performs a number
of cycles about the Sun, repeatably passing near to the Earth. In a time span 
roughly on the order of 100 years it collides with the Earth on a near parabolic
orbit.
\medskip
\medskip

We present here a mechanism for the origin of a Mars-sized Earth
impactor and describe the path it would take to arrive at Earth
collision via a special class of slowly moving chaotic collision 
trajectories. The analysis shows that Earth 
collision along these trajectories is likely. 
Approaches for further work are discussed in the Appendix.
\medskip
\medskip
\medskip

\noindent
{\it Note added in proof}
\medskip

        As we have discussed, the giant impactor could have grown
up in a stable orbit at Earth's $L_4$ (or $L_5$) point where a stable orbit is possible 
and an object could remain and be able to grow by accretion without hitting the 
Earth early-on. We expect this phenomenon could occur when there was a thin disk 
of planetesimals (in nearly circular orbits). We have noted that Saturn's rings 
are an example of such a thin disk of planetesimals(in this case, chunks of ice
plus some dirt) observable today. Saturn's regular icy moons (inside the orbit of
Titan) are all in nearly circular orbits of low eccentricity suggesting that they formed
out of a thin disk of planetsimals(ice chunks) rather like Saturn's rings today only
larger in extent. In such a situation we might expect our scenario to operate.
Therefore 
it is quite interesting that we can find examples of objects at $L_4$ (or $L_5$), or escaping 
from $L_4$ (or $L_5$) in the Saturn system. Saturn's moon Helene co-orbits at 
the $L_5$ point (60$^o$ ahead) of the larger moon Dione. Helene has a largest 
diameter of 36 km and Dione has a diameter of 1120 km. Saturn's moons Telesto 
(diameter 34 km) and Calypso (diameter 34 km) occupy both the $L_4$ and $L_5$ points 
relative to Saturn's moon Tethys (diameter 1060 km). We would say that Helene, 
Telesto and Calypso originated in a planetesimal disk(of ice chunks) at these stable Lagrange 
points and have grown in place there surviving till the present without colliding 
with Dione or Tethys. The rest 
of the planetesimals(ice chunks) have accreted onto the regular moons of Saturn.(Saturn's
rings themselves lie inside the Roche limit where the formation of large objects
is forbidden by accretion.) While 
these Lagrange moons are small relative to the primary, growth of larger objects
with respect to the primary is also possible. Saturn's moons Epimetheus (119 km
diameter) and Janus (179 km diameter) co-orbit in horseshoe orbits just like the 
one we found for the giant impactor near breakout (Figure 5).  We would say that 
Epimetheus formed at a Lagrange point of Janus and grew along with it by accretion 
from the planetesimal disk. Later perturbations by other planetesimals kicked it 
out into a horseshoe orbit just short of breakout. Thus, an object (Epimetheus) 
nearly as large as the primary (in this case Janus) can form and end up 
in a horseshoe orbit. Just a little more perturbation and Epimetheus would achieve 
breakout and likely collide with Janus.  These provide examples of the phenomena 
described in this paper that can be observed today.    

        A similar pair of co-orbiting objects in horseshoe orbits in another 
solar system could be easily detected using stellar radial velocity data. This would 
appear to be a planet in circular orbit about the star whose mass was observed
to mysteriously vary. The mass variation would be approximately sinusoidally in
time with a period significantly longer than the orbital period of the primary. 
For example, if the secondary had a mass 0.1 times that of the primary (like the 
giant impactor) then this would show up as a nearly sinusoidal variation of 10$\%$ 
in the deduced mass of the primary. If the two were nearly equal in mass there 
would be a 100$\%$ variation in the mass.  (When they were near each other at one
end of the horseshoe orbit the effective mass perturbing the star would be nearly 
doubled, and when they circulated to be on opposite sides of the star their perturbation 
would temporarily vanish.)  We should have a look among the known cases
of extra-solar planets for such cases. Granted, we are currently able to see only gas 
giant planets(which may have even migrated inward) rather the terrestrial ones we are considering, but still 
it would be interesting to look. If one found such a case, it would be easy to 
prove. 

      Also of particular interest is the Earth co-orbiting asteroid 2002 $AA_{29}$ 
which is in a horseshoe orbit relative to the Earth. Of course, in addition to the 
giant impactor there can be other Lagrange-point debris particularly at the other
stable Lagrange point not occupied by the giant impactor. This material may have 
been kicked out early-on by other planetesimals or by the giant impactor itself 
as it escaped into a horseshoe orbit. Does any of this material survive to the 
present day? 2002 $AA_{29}$ (diameter $<$ 0.1 km) is in a horseshoe orbit at 1AU, 
virtually identical to the horseshoe orbits found
by us in Figure 5. This asteroid approaches the Earth closely (3.6 million miles away) 
once every 95 years while circling the sun at 1AU. It was near one of 
these close approaches that it was discovered in 2002. After a number of cycles, 
it is briefly captured for a period of 50 years as a quasi-satellite of the Earth, 
before returning to the 95-year horseshoe orbit cycle. Its rather large 
inclination (10.7$^o$) saves it from collision with the Earth. This object may have 
originated near $L_4$(or $L_5$), and have been kicked out into a horseshoe orbit (perhaps 
by the giant impactor itself). If that is so, it could be composed of the 
same material that also formed the seeds for the Earth and the giant impactor. A
sample return from this asteroid thus offers the possibility of obtaining some 
primordial material from the same reservoir that produced the Earth and the Moon. 
In this case, it should have oxygen isotope abundances similar to those found 
for the Earth and the Moon and an iron abundance similar to that of the Earth. The final 
oxygen isotope abundances and iron abundances of the Earth reflect not 
only their seed material (originally from 1AU) but also the integral of the abundances 
accreted later, from material originally inside and outside 1AU. Thus, any 
slight differences in oxygen isotope abundances would be helpful in illuminating 
the accretion process. It would of course be very interesting to measure the 
age of a 2002 $AA_{29}$ sample. On the other hand, if the sample has oxygen isotope
abundances identical with the Earth and the Moon, but is poor in iron like the 
Moon, that would suggest it was part of the splash material kicked out by the 
giant impactor at near escape velocity which did not coalesce onto the Moon but rather 
ended circling the Sun at 1AU and then became trapped at $L_4$(or $L_5$) where it moved for 
perhaps a considerable time before finally being kicked out. If 
the sample has completely different oxygen isotope abundances from those of the 
Earth and Moon, that would indicate an origin elsewhere in the solar nebula (not
at 1AU) and we would then have to explain how it somehow got perturbed into a low 
eccentricity horseshoe orbit at 1AU. (Most Earth-crossing asteroids perturbed
into their current orbits from the main belt should have much larger 
eccentricities according to Ipatov and Mather 2002.) Bottke, et al (1996) have previously 
suggested that low eccentricity objects near 1AU could have an origin tracing 
back to the Earth-Moon system, and radar results suggest (Ostro, et al 2003) that 
2002 $AA_{29}$ has a high albedo which supports this hypothesis (according to Connors
et al 2004).  A sample return from asteroid 2002 $AA_{29}$ is thus of particular scientific interest 
and may provide important clues as to the origin of the Earth and 
the giant impactor that formed the Moon.  
      
\medskip

\noindent
{\bf Acknowledgments}
\medskip

\noindent
We would like to thank Scott Tremaine and Peter Goldreich for helpful comments, 
and also Robert Vanderbei for use of his solar system simulator. 
\medskip

\noindent 
Partial
support for this work for J. Richard Gott, III is from NSF grant AST-0406713, and 
for Edward Belbruno from grants by NASA, Office of Space Science, and
Goddard Space Flight Center.

\newpage

\appendix{\bf Appendix: Planetesimal Dynamics in a Thin Disk and Random
Encounters with the Great Impactor and Earth}
%\section{Appendix}
%\label{ref:Appendix}

\medskip

\noindent
%{\bf Planetesimal Dynamics in a Thin Disk and Random
%Encounters with the Great Impactor and Earth}
\medskip

We estimate the magnitudes of the velocity perturbations on the giant 
impactor and the Earth due to encounters with other small 
planetesimals with a back of the envelope
calculation. Such velocity perturbations 
drive the impactor of mass $M_I$ into breakout. 

Let $\mu$ be the typical mass of the remaining planetesimals (or more 
precisely the rms mass observed in the distribution)(Note: This definition 
of $\mu$ is different than the definition of $\mu$ in the discussion of the three-body
problem, equation \ref{eq:2}).  Consider the disk
of planetesimals to be of thickness
$$
\sim2(V_{\mu}/V_{orb})(1AU)\sim2r_E(V_{\mu}/4\times10^{-5}V_{orb})
$$ 
where $V_{\mu}$ are the typical peculiar velocities of the planetesimals.
and include all radii nearer to 1AU than either Venus or Mars. The
volume of the disk is then $2(V_{\mu}/V_{orb})\pi(.85)(1AU)^3$. Let the
planetesimals have peculiar velocities of order $V_{\mu}$, or
equivalently orbital eccentricities $e$(and inclinations $i$) of order 
$V_{\mu}/V_{orb} \sim .0006$, where $V_{orb} = 29.86$ km/s. 
The thickness of the disk is then of order 30$r_E$.
For distant encounters, with impact parameter $b$, the deflection of the
planetesimal by the impactor is
$$
\delta V_{\mu}/V_{\mu} \sim 
\left\{
\begin{array}{cc}
2GM_I/V_{\mu}^{2}b &\mbox{ if } b > b_c, \\
1                  &\mbox{ if } b \leq b_c , \\
\end{array}
\right.
$$
where $b_{c}=2GM_I/V_{\mu}^{2}$, and $m_I$ is the mass of the giant impactor.

These are hyperbolic encounters, and the planetesimal exits the
encounter with the same magnitude of peculiar velocity $V_{\mu}$ but
changed in vector direction.  Thus, for nearby encounters, $\delta
V_{\mu}/V_{\mu}$ can be at most 2. Momentum is conserved so the kick in
velocity $\delta V$ received by the giant impactor is given by
$$ \delta V \sim\delta V_{\mu} \mu /M_I, $$ 
so for a single collision
$$
(\delta V)^2 \sim
\left\{
\begin{array}{cc}
4G^2{\mu}^2/V^2_{\mu}b^2   &\mbox{ if } b > b_c, \\
V_{\mu}^{2}{\mu}^2/M^2_I   &\mbox{ if } b \leq b_c . \\
\end{array}
\right.
$$

The collisions are independent so the velocity kicks add in quadrature,
giving a random walk in velocity space with time. Since the disk is thin
with $V_{\mu}/V_{orb} \sim .0006,$ the half-thickness of the disk is
.0006AU $\sim 14 r_E$, most of the time $ b \gg 14 r_E$ and the
encounters are mostly in the plane(i.e. $0 < b_{\perp} < 14 r_E$, while
$0 < b_{\|} < 1AU$) and 
$$ (\delta V_{\perp})^2/(\delta (V_{\|})^2 \sim b^2_{\perp}/b^2_{\|},$$
implying that the major component of the $\delta V$ is parallel to the
plane. The number of planetesimals is $n = M_{disk}/{\mu}$ so
considering the geometry of the disk (seen edge on it is a horizontal
strip of thickness $2(V_{\mu}/V_{orb})1AU)$), the number of collisions
with impact parameter $b$ ($b > 14 r_E$) from that strip within a time t
is:
$$ \delta N  \sim n \left[2(V_{\mu}/V_{orb})\pi
(.85)(1AU)^3 \right]^{-1} 4(V_{\mu}/V_{orb})(1AU)db V_{\mu}t.$$

The relevant area is $4v(1AU)db$, where $v \equiv V_{\mu}/V_{orb}$,
because the disk has thickness of $2v(1AU)$ and there are two vertical
strips of width $db$ and height $2v(1AU)$ at a distance $b$ from the
giant impactor(one to the left and one to the right). The range of the
impact parameters we should integrate over is approximately 0 to 1AU.

The impact parameter for physical impact is calculated from the 
parameters of the hyperbolic
encounter. When the small planetesimal hits $M_I$ it will have a velocity
$V_S$, where $.5 V_S^2 = .5 V_{\mu}^2 + GM_I /r_I$. At the maximum
impact parameter for physical collision the planetesimal will just graze $M_I$ 
at the perigee of its orbit, so that $V_S$ will be tangential at that
point and its angular momentum per unit mass will be $L = V_S r_I =
V_{\mu} b_m$ (or equal to what it had initially). Thus, $b_m = 
r_I V_S / V_{\mu}$ and
$$
b_m \sim r_{I}(1+[ 2GM_I /r_I V_{\mu}^2])^{1/2} \sim
r_I (1 + [b_c /r_I])^{1/2} \sim b_c ([r_I^2 /b_c^2 ] +
[r_I /b_c ])^{1/2},
$$ and since the escape velocity squared $V_{es}^2$ from the surface of the giant impactor is much
greater than $V_{\mu}^2$, then $r_I /b_c = V_{\mu}^2 /V_{es}^2 \ll 1$ and
$b_m \sim (b_c r_I)^{1/2}$. (We are considering here only gravitational
encounters, not physical collisions, so $b_m$ will be the minimum impact parameter for our
integration. Of course, there will be some contribution to 
$(\delta V_{\|})^2$ from direct collisions and accretion but this is difficult to
calculate and we will ignore this contribution in this simple treatment.
Since $(b_c r_I)^{1/2}$ is small relative to 1AU, this contribution to $(\delta V)^2$ 
will be negligible in any case. The effects of direct collisions on
$(\delta V_{\perp})^2$ may be significant, but again difficult to
calculate, so for simplicity we are ignoring them and only considering
gravitational encounters.)

Thus, we will integrate from $(b_c r_I)^{1/2}$ to 1AU, and in the
regime we are interested in $v \sim .0006, (b_c r_I)^{1/2} < 1AU$, and
$b_c > 1AU$, so we are in the regime where $(b_c r_I)^{1/2} < b < 1AU$ and:
\begin{equation}
(\delta V_{\|})^2 \sim (M_{disk}/{\mu})[2v\pi (.85)(1AU)^3]^{-1}4v(1AU)
V_{\mu}t\int^{1AU}_{\sqrt{b_c r_I}} [V^2_{\mu} \mu^2 /M^2_I] db  .
\label{eq:A2}
\end{equation}
This yields,
$$
(\delta V_{\|})^2/V^2_{orb} \sim [(M_{disk}/{.085\mu})]4v^3
(t/yr)(\mu/M_I)^2.
$$
Breakout is achieved when $(\delta V_{\|})^2 \sim (.006)^2 V_{orb}^2$ or
after a time
$$
t_{break} \sim 7.65 \times 10^{-6} v^{-3} (\mu / M_{disk})(M_I /{\mu})^2 yrs.
$$
Since $v = 0.0006$, then
\begin{equation}
t_{break} \sim 35,400 yrs (\mu /M_{disk}) (M_I / \mu)^2.
\label{eq:AA2}
\end{equation}

At breakout we might expect $M_{disk} \sim 0.3 M_I$ since we want
subsequent accretion onto the Earth to be inconsequential relative to
$M_I$. Canup and Asphaug's successful simulation has the giant impactor
hit a nearly formed Earth. For example with $\mu \sim M_I/300$ 
we expect $n \sim 100$ other large planetesimals of mass 
$\mu \sim M_I/300$ and $t_{break} \sim 35$ million years.
Long enough for iron cores to form in the Earth and the giant
impactor as required and in agreement with radioactive halfnium-tungsten
chronometer results (see Canup (2004)).

The same calculation as done leading to equation \ref{eq:A2} above could be repeated with $M_E$
replacing $M_I$ and one would see that
\begin{equation}
(\delta V_{\|})^2_E /(\delta V_{\|})^2_I \sim M_I^2 /M_E^2.
\label{eq:A1}
\end{equation}
Thus the rms peculiar velocity acquired by the Earth due to velocity
kicks in time t is 1/10th as large as that acquired by the giant
impactor.

Returning to the giant impactor we see that for individual collisions
$$ (\delta V_{\perp})^2 \sim (\delta V_{\|})^2 (b_{\perp}^2 /b_{\|}^2),$$
where
$$
<b_{\perp}^2> \sim \int^{V_{\mu}(1AU)/V_{orb}}_0 x^2 dx / 
\int^{V_{\mu}(1AU)/V_{orb}}_0 dx \sim (1/3)v^2(1AU)^2 . 
$$
Similar to the estimation of $(\delta V_{\|})^2$ in (\ref{eq:A2}), it is found that
$$
(\delta V_{\perp})^2 \sim [ M_{disk}/(0.85{\mu}(1AU)^2) ]2V_{\mu}t(1/3)v^2(1AU)^2
\int^{1AU}_{\sqrt{b_c r_I}} [V_{\mu}^2 {\mu}^2 /(b^2 M_I^2)] db
$$ which simplifies to
\begin{equation}
(\delta V_{\perp})^2 \sim [(M_{disk}/(0.85{\mu})](2/3)V_{\mu}tv^2
[V^2_{\mu}/(b_c r_I)^{1/2}M_I^2].
\label{eq:A3}
\end{equation}
Equations (\ref{eq:A2}), (\ref{eq:A3}) imply
\begin{equation} 
(\delta V_{\perp})^2/(\delta V_{\|})^2 \sim (1/3)v^2[1AU/(b_c
r_I)^{1/2}].
\label{eq:A4}
\end{equation}
Now $(b_c r_I)^{1/2} \sim r_I(V_{es}/V_{\mu})$, and it can be shown that
(\ref{eq:A4}) reduces to
$$
(\delta V_{\perp})^2/(\delta V_{\|})^2 \sim  (1/3)(V_{\mu} v^2
/V_{es})[1AU/r_I].
$$
If $V_{\mu} \sim .0006 V_{orb}$, 
$$
(\delta V_{\perp})^2/(\delta V_{\|})^2 \sim 1.9 \times 10^{-5}.
$$
So when the giant impactor achieves breakout $\delta V_{\|} \sim .006 V_{orb},
\delta V_{\perp} \sim .000026 V_{orb}$, so if we simulate the random walk by
applying random kicks 
\begin{equation}
\delta V_{\|}/V_{orb} \sim .001, \ \ \ \ \delta V_{\perp}/V_{orb} \sim .0000044 
\label{eq:A7}
\end{equation}
to the giant impactor at random times, after approximately 36 kicks
breakout should be achieved. If we followed this to its conclusion that
would mean an inclination at breakout for the giant impactor of
$i \sim \delta V_{\perp}/V_{orb} \sim .000026$ radians $\sim$ 5 sec
$\sim$ .6$ r_E /(1AU)$, which would keep it on an easy collision course
with the Earth. This also guarantees that the collision will be nearly
in the plane of the ecliptic (i.e. $b_{\perp}/b_{\|} \sim 0.6 r_E/0.006 AU \sim 4 \times 10^{-3}$
or within 0.25$^o$ of the ecliptic). This should, by consideration of the total
angular momentum, produce an orbit for the Moon approximately in the plane
of the ecliptic as is observed. Some debris is ejected so the alignment should just
be approximate -which it is. Interaction of the debris disk with the Earth as the Moon 
forms can also naturally lead to a moderate tilt of the Earth relative to the Moon's
orbit as explained by Canup $\&$ Asphaug (2001).

The calculation for $(\delta V_{\perp})^2$ can be repeated for the
Earth, yielding
$$
(\delta V_{\perp})^2_E/(\delta V_{\|})^2_E \sim 
(1/3)(.0006)^3 (V_{orb}/V_{Ees})[1AU/r_E ] \sim 4.5 \times 10^{-6},
$$
where $b_c = 2GM_E /V_{\mu}^2$, since $r_E /b_c = V^2_{\mu} /V^2_{Ees} =
(.018$ km s$^{-1}$/11.19 km s$^{-1})^2 = 2.6 \times 10^{-6}, b_c = 
16.4 AU.$ 

Thus, by the time the giant impactor has achieved breakout the values 
for the peculiar velocity of the Earth will be:
\begin{equation}
(\delta V_{\|})_E /V_{orb} \sim .0006, \ \ \ \ (\delta V_{\perp})_E /V_{orb}
\sim .0000013.
\label{eq:A5}
\end{equation}

Therefore, we will apply velocity kicks of 
\begin{equation}
(\delta V_{\|})_E /V_{orb} \sim .0001, \ \ \ \ (\delta V_{\perp})_E
/V_{orb}
\sim .0000002.
\label{eq:A6}
\end{equation}
at random times to the Earth and after 36 kicks they should achieve by
random walk the values given by (\ref{eq:A5}). As can be seen, the
movement of the Earth is less than that of the giant impactor, so one
may say that although the motion of the Earth complicates the situation
it is still basically the giant impactor that is achieving breakout. The
Earth is basically a spectator while the giant impactor breaks out of
its stable equilibrium at $L_4$.

Let us calculate the minimum impact parameter $b_m$ for collision
between $M_I$ and $M_E$. This is a two-body problem. Impact occurs
when the separation of the centers of $M_I$ and $M_E$ is $\rho = 
r_I + r_E$. $M_I$ has a peculiar velocity relative to $M_E$ of order
$V_{break}$. Imagine shooting $M_I$ at the Earth with relative velocity
$V_{break}$ and a minimum impact parameter $b_m$ so that is just has a
grazing collision with the Earth. At that point of collision the relative velocity is
$V_S$ and the angular momentum per unit mass is $V_S \rho = V_{break}
b_m$, which is the initial angular momentum per unit mass. Now,
conservation of energy gives
$$
{1\over2} V_S^2 = {1\over2} V_{break}^2 + [G(M_I + M_E )/\rho]
$$
and this yields
\begin{equation}
b_m = \rho(1+[(V_{es}^2 /V_{break}^2)(M_I +M_E )r_E /\{ M_E \rho
\}])^{1/2},
\label{eq:bm}
\end{equation}
where $V_{break} = .006 V_{orb} = .18$km s$^{-1}$, $V_{es} = 11.19$ km
s$^{-1}$ = $62.1 V_{break}$ is the escape velocity from the surface of
the Earth. Now, $(M_I + M_E )/M_E = 1.1$, and $\rho /r_E = 1.53 r_E $.
Thus, (\ref{eq:bm}) yields
$$
b_m = 1.53 r_E (1 + [3856.4/1.39])^{1/2} = 80.6 r_E = .0034 AU.
$$
Now, if the vertical peculiar velocity obtained by the giant impactor
upon breakout is less than .0034 $V_{orb}$, as it is in the thin disk
case we are considering, then it will not be missing the Earth in the
vertical direction. It will have a typical eccentricity of .006 and will
therefore have a typical impact parameter with respect to the Earth in the plane of
order .006 AU. So the chance of impact in the thin disk case on the
first pass by the Earth is of order
$$
\mathcal{P} \sim (b_m /.006AU) \sim .57
$$
or a substantial probability. In fact we observe $\mathcal{P} \sim .25$
as discussed in Step 6 of Section 3. In other words we had one success 
on hitting the Earth on the first pass on four
attempts. Our earlier tests with the restricted three-body problem in
the plane had 4 successes out of 13 attempts for $\mathcal{P} \sim .3$,
as we discussed at the end of Step 2 in Section 3 and given by
(\ref{eq:prob1}). Interestingly, there we picked four directions
$+x,+y,-x,-y$, where the values of $V_{break} /V_{orb}$ were
respectively, .007,.011,.007,.012. The above calculation indicates that
roughly $\mathcal{P} \propto (1/V_{break})$, so given the values of
$V_{break}$ we would expect $\mathcal{P} \sim .49$ for the $\pm x$
directions where $V_{break} = .007$, and, in fact, we had 5 runs with
two successes at hitting the Earth within 647 years so we observe 
$\mathcal{P} = .4$. By comparison, we would predict $\mathcal{P} \sim
.30$ for the $\pm y$ cases where $V_{break} = .0115$. In fact, in those
cases we had  2 successes in hitting the Earth in 647 years out of 8
runs giving $\mathcal{P} = .25$. In both the $\pm x$ and $\pm y$ cases
the results are comparable with our estimates. As discussed previously, these 
are all rough numbers which could be improved by doing additional trials.

(In the case of a thick disk, the perturbations on the giant impactor
would be of order $i \sim .006 = V_{break} / V_{orb}$ and therefore it
would have a probability of missing in the vertical direction of order
$(b_m /.006)AU$, as well as a similar probability of missing in the plane
so the chance of impacting the Earth on the first pass would be of order
$\mathcal{P} \sim [(b_m /.006)AU]^2 \sim .32$, which is smaller but
still appreciable.)

As we have derived above in equation \ref{eq:AA2} 
$$
t_{break} \sim 35,400 years (M_I/M_{disk})(M_I/\mu)
$$

For $M_{disk} \sim 0.3 M_I$ and $\mu \sim M_I/300$, this gives a
breakout time of $\sim 35$ million years. 

Since the planetesimals are in orbits with eccentricities $e \sim
V_\mu/V_{orb}$ and the accretion impact parameter $b_a >
1AU(V_\mu/V_{orb})$, the only planetesimals that can hit the great
impactor or the Earth are those at a radius from the Sun of 
$1AU - b_{aI} < r < 1AU + b_{aI}$ and $ 1AU - b_{aE} < r < 1AU + b_{aE}$,
respectively. (Recall that the impact parameter for accretion for a
planetesimal onto the giant impactor is $b_{aI} \sim r_I(1 +
V^2_{es}/V^2_{\mu})^{1/2} \sim 270 r_I \sim 9.1 \times 10^5$ km, and
onto the Earth is $b_{aE} \sim r_E(1 +
V^2_{es}/V^2_{\mu})^{1/2} \sim 622 r_E \sim 4.0 \times 10^6$ km $>
V_\mu/V_{orb} 1AU = 8.9 \times 10^4$ km.) But once a band of width
$2b_{aE}$ is cleared out there will be no further prompt physical
collisions with the Earth or the giant impactor.

So the Earth and the giant impactor will quickly clear out that area,
but further accretion will await scattering of planetesimals into that
region on a timescale dictated by two-body relaxation among the
planetesimals. The planetesimal disk closer to 1AU than to Venus or Mars
has limits 0.86AU to 1.26AU. So, to accrete the entire remaining disk, a timescale
is required similar to that for a planetesimal to random walk up to an
eccentricity of order 0.2, or equivalently to acquire a peculiar velocity
in the disk of order $(\delta V{\|}) \sim (0.2)V_{orb}$. Let us estimate
this accretion timescale. 

Let $\mu$ be the typical mass of the remaining planetesimals (or more
precisely the rms mass observed in the distribution). Let the peculiar
velocities of the planetesimals be $V_{\mu} \sim 0.0006 V_{orb}$. The
disk of the planetesimals is to be of thickness $2 \times 0.0006 (1 AU)$
and includes all radii nearer to 1AU than either Venus or Mars. The
volume of the disk is then $2(V_\mu/V_{orb})\pi(0.85)(1AU)^3$. For
distant encounters, with impact parameter $b$, the deflection of the
planetesimal upon passing another planetesimal is
$$
\delta V_{\mu}/V_{\mu} \sim 
\left\{
\begin{array}{cc}
2^{1/2}G\mu /V_{\mu}^2 b & \mbox{ if } b > b_c = 2^{1/2}G\mu /V_{\mu}^2, \\
1                        & \mbox{ if } b \leq b_c, \\
\end{array}
\right.
$$
where we take into account the fact that the total mass of the two
particle system is $2\mu$, the rms relative velocity between the two 
particles is $2^{1/2}V_{\mu}$, and $\delta V_{\mu}$ is 1/2 of the total
change in relative velocity. 

So for a single collision:
$$(\delta V_{\mu})^2 \sim
\left\{
\begin{array}{cc}
2 G^2 \mu^2  /V_{\mu}^2 b^2 & \mbox{ if } b > b_c, \\
\\
V_{\mu}^2                    & \mbox{ if } b < b_c. \\
\end{array}
\right.
$$

The collisions are independent so the velocity kicks add in quadrature, giving a
random walk in velocity space and time. Since the disk is thin,  
$b_c \gg (V_{\mu}/V_{orb})1AU$,  
$(0 < b_{\perp} < (V_{\mu}/V_{orb})1AU$), while $0 < b_{\|} < 1AU$ so the encounters are mostly in
the plane and the major
component of $\delta V_{\mu}$ is parallel to the plane. The number of planetesimals
is $M_{disk}/\mu$ so considering the geometry of the disk [seen edge on it is a
horizontal strip of thickness $2(V_{\mu}/V_{orb})1AU$], the number of collisions with
impact parameter $b$ (where $b > (V_{\mu}/V_{orb})1AU$) from that strip within a time
$t$ is:
$$
\delta N \sim
(M_{disk}/\mu)[2(V_{\mu}/V_{orb})\pi(0.85)
(1AU)^3]^{-1}4(V_{\mu}/V_{orb})1AU2^{1/2}V_{\mu}t db.
$$
The range of the impact parameters $b$ we should integrate over 
is approximately $b_m$ to 1AU. The rms impact velocity is $2^{1/2} V_{\mu}$.
\noindent
Considering the relative velocities of the two planetesimals and that physical impact
occurs when their center-to-center separation is $2 r_{\mu}$, the minimal impact
parameter for the physical impact is
$$
b_m \sim 2r_{\mu}(1+G \mu /r_{\mu}V_{\mu}^2)^{1/2} \sim 2r_{\mu}(1 +
(b_{c}/2^{1/2})r_{\mu})^{1/2}.
$$
Now, 
$ r_{\mu} \sim (\mu/M_I)^{1/3} 3380$ km, and $b_c \sim 2^{1/2} G \mu /V^2_{\mu} \sim
1.75 \times 10^8 km (\mu/M_I)(0.0006 V_{orb}/V_{\mu})^2$. So $b_c/r_{\mu} \sim 5.18 \times
10^4 (\mu / M_I)^{2/3}(0.0006 V_{orb}/V{\mu})^2$, and if the remaining planetesimals
are large but still smaller than the giant impactor (say $\mu / M_I \sim 1/300$) then
$b_c /r_{\mu} \gg 1, b_m \sim 2^{3/4}(b_c r_{\mu})^{1/2}$, and $1AU \gg b_c \gg b_m
> (V_{\mu}/V_{orb} 1AU \gg r_{\mu}$.

(We are considering here only gravitational encounters, so $b_m$ will be the minimum
impact parameter for our integration. Of course there will be some contribution to
$(\delta V_{\|})^2$ from direct collisions which is difficult to calculate and we will
ignore this contribution in this simple treatment. Since $b_m$ is small relative to
$b_c$, this contribution to $(\delta V_{\mu})^2$ will be negligible in any case.)

\noindent
Thus, we will integrate from $b_m = 2^{3/4}(b_c
r_{\mu})^{1/2}$ to 1AU:
$$
(\delta V_{\|})^2 \sim
(M_{disk}/\mu)[2 \pi(0.85)(1AU)^2]^{-1}(4)2^{1/2}V_{\mu}t\{\int^{b_c}_{b_m}
V_\mu^2 db + \int^{1AU}_{b_c} 2G^2 {\mu}^2 / V_{\mu}^2
b^2 db\}.
$$
This yields,
$$
(\delta V_{\|})^2 \sim
(M_{disk}/{\mu})[(0.85)(1AU)]^{-1}(4)2^{1/2}(V_{\mu}/V_{orb})(t/yr)\{[b_c
(V_{\mu}^2)]+[2G^2{\mu}^2 /V_{\mu}^2 b_c]\},
$$
which reduces to,
$$
(\delta V_{\|})^2 \sim 
(M_{disk}/{\mu})[(0.85)(1AU)]^{-1}16 (V_{\mu}/V_{orb}) (t/yr) [G\mu],
$$
yielding,
$$(\delta V_{\|})^2 / V_{orb}^2 \sim
[M_{disk}/(0.85)M_{Sun}]16 (V_{\mu}/V_{orb})(t/yr).
$$
Thus, the accretion timescale is of order
$$
t_a \sim (0.2)^2 [(0.85)M_{Sun}/16M_{disk}](V_{orb}/V_{\mu})yr,
$$
which yields $t_a \sim 11.8$ million yrs $[M_I/M_{disk}]$.

Now, recall that the breakout timescale is
$t_{break} \sim$ 35,400 years $(M_I/M_{disk})(M_I/{\mu})$.

If we want $t_a > t_{break}$ then 
$$
\mu/M_I \stackrel{>}{\sim} 1/300.
$$

Thus, we expect that the remaining planetesimals would have had time to
grow large, but would still be smaller than the giant impactor. If
$M_{disk} \sim 0.3 M_I$, and $\mu \sim M_I /300$, there would be of
order 100 large ($\sim$500 km in radius) planetesimals, and the breakout
timescale for the giant impactor would be of order 35 million years 
while the timescale for accretion of the remaining large planetesimals 
would be of order 40 million years. This gives enough time for the Earth 
and the giant impactor to form and for their iron cores to sink into their
centers (in agreement with the estimate of 10-30 million years for core
formation from the radioactive halfnium-tungsten chronometer (see
discussion in Canup (2004)).

So if we want the breakout to occur on a timescale shorter than the
remaining accretion timescale, we would want $\mu/M_I \sim 1/300$ or
greater. Thus, we expect by the time the giant impactor is achieving
breakout, there would be just a few($<$100) large planetesimals 
left dominating
the mass distribution. Reasonable, since the largest of the remaining
planetesimals left at the time of breakout would have had time to 
grow large. Still we would expect the giant impactor to be the largest of
the remaining planetesimals.
 
After the giant impactor hits the Earth, and the Moon is formed from
the
splash debris, then the Earth will continue to accrete the remaining
planetesimals. The accretion cross section for the Earth is of order
$\sigma_E \sim \pi(b_c R_E) \sim \pi r_E^2 (V_{Ees} / V_{\mu})^2 \sim
\pi (62r_E)^2$. In Canup and Asphaugh's model the Moon is expected to
form at $1.2 a_{roche} = 3.5 r_E$ (much later it drifts out to its
current location by tidal interaction). The cross section for bringing
an object inside the Moon's orbit radius is $\sigma_{EM} \sim \pi(3.5
r_E)^2 (V_{Ees}^2/3.5V_{\mu}^2) \sim 3.5 \sigma_E$. Of those crossing
the Moon's orbit, only 1/3.5 will hit the Earth on that pass. Of those
crossing the Moon's orbit, at the time they cross they will have a
velocity of $V^2 \sim V_{\mu}^2 + V_{Ees}^2/3.5 \sim V_{Ees}^2/3.5$,
and at
this velocity the Moon's accretion cross section is $\sigma_m \sim
\pi r_M^2 (1+ [3.5 V_{Mes}^2/V^2_{Ees}])^{1/2} \sim 1.16 \pi r_M^2$.
The fraction that cross the Moon's orbit that impact the Moon will
therefore be $ f \sim 1.16 \pi r_M^2 / 4\pi(3.5r_E)^2 \sim 1.7 \times
10^{-3}$. So the number of objects hitting the Earth is larger than the number hitting
the Moon by a factor of $(1/3.5)/1.7 \times 10^{-3} \sim 160$.  
If less than 160 large objects are eventually
accreted then the average number expected to hit the Moon is less than
1. So
if the number of large objects accreting is $<$100,
there is an appreciable chance that all those large objects will
hit the Earth and none will hit the Moon.
This an important advantage for our model since it does
not pollute the Moon with any additional iron - leaving it iron poor.
Indeed, this reason is cited by Canup $\&$ Asphaugh (2001) in
arguing that accretion on the Earth and Moon after the great impact be
small. The Moon may be expected not to gain appreciable additional
material. 
Still smaller planetesimals, which make an insignificant contribution to
the total remaining disk mass, may fall on the Earth and Moon-creating
impact sites like Mare Imprium without adding significantly to the mass.
Since our scenario depends on the fact that there will be some debris
left in Earth's neighborhood at the time of the formation of the Moon by
the giant impactor, it is a plus that the Moon shows some signs of
late impacts itself.

The parameters here can be considered as a toy model at best. We have
ignored dynamical friction.
Dynamical friction could
slow breakout by slowing the accumulation of peculiar velocity of a massive body, but since
$L_4$ is at a peak of the effective potential dynamical friction might even speed breakout.
We have also ignored the effects of momentum transfer
perpendicular to the plane in planetesimals that actually hit the giant
impactor(considering only the momentum transfer in the more frequent
distant encounters), because the latter is difficult to calculate.
We have considered peculiar and parallel velocities as if they occurred
in a slab ignoring the Keplerian motion around the Sun. We need not 
be married to these particular parameters, as they just form a jumping
off point. They give us a guess as to the ratio of perpendicular
velocities to velocities in the plane that might be acquired by
gravitational perturbations in a random walk scenario. The random walk
scenario is one that should occur under very general circumstances.
Other model parameters and assumptions might lead to varying scenarios,
but those we have shown are a starting point for discussion. A natural
continuation would be N-body experiments.
More complicated mass distributions and various eccentricity models for
the planetesimals could be considered. There is a lot of parameter space
to be explored. More
elaborate simulations with millions of particles including treatment of 
physical collisions could simulate the
formation of the Earth and the giant impactor and their growth in a cold
disk scenario. After gaseous dissipation was finished and only
planetesimals were left, we expect some debris to remain at $L_4$ and
$L_5$ (like the Trojan asteroids) because planetesimals either in, or perturbed,
stable orbits about about $L_4$ and $L_5$ would stay there. A large
object like a giant impactor can grow by accretion at $L_4$(or $L_5$).
It is a matter of survival: an object in a stable orbit at $L_4$(or
$L_5$) will survive--not hitting the Earth--and by surviving can have
more time to accrete other planetesimals and grow large itself. 
One could see how often the second largest object growing near 1AU in
fact started in the Lagrange point debris of the Earth. And finally, in
the cases where a giant impactor of the type required to form the Moon
did indeed hit the Earth causing formation of a moon like ours, one
could see how often that impactor did in fact originate in Lagrange
point debris.
In other words, what is the probability that the giant impactor originated at $L_4, L_5$ 
{\it given} that a moon like ours(with material of identical oxygen abundance from 1AU)
is formed by collision?

There are several races going on. The Earth starts forming by accretion
and as it grows in mass, a stable Lagrange point at $L_4$(and $L_5$)
forms. As we have shown, as the Earth grows, the region of stable
orbits around $L_4$ grows in size, so a planetesimal
trapped there in a stable orbit would stay there as a proto-Earth grew.
It would grow larger itself by surviving and accreting other
planetesimals. It must grow to a mass of order $0.1M_E$, by the time it
is perturbed out of its stable orbit and achieves breakout. Breakout
must be achieved before all the remaining planetesimals have accreted
onto the Earth. After breakout, a collision with Earth on a near parabolic
trajectory is likely.

Our paper points out the possibility that the giant impactor that formed the Moon
could have originated at $L_4$, survived there long enough to grow large by accretion, and
eventually been perturbed by other planetesimals onto a collision course with the Earth.
Awaiting numerical simulations capable of showing this occurring in detail, we have used examples
within our own solar system to support our model (see note added).

\bigskip\bigskip\bigskip\bigskip\bigskip\bigskip\bigskip\bigskip\bigskip\bigskip\bigskip

\end{document}